\documentclass[twocolumn, 10pt]{IEEEtran}
\usepackage[cmex10]{amsmath}
\usepackage{graphicx,amssymb,amsfonts,bm,setspace}
\usepackage[noadjust]{cite}
\usepackage{braket} 		
\usepackage{amsthm}

\allowdisplaybreaks

\newtheorem{cor}{Corollary}
\newtheorem{defn}{Definition}

\newtheorem{lem}{Lemma}

\newtheorem{thm}{Theorem}

\newcommand{\abs}[1]{\left| #1 \right|}

\newcommand{\expect}[1]{\mathbb{E}\left[#1\right]}

\newcommand{\prob}[1]{\Pr\left[#1\right]}

\newcommand{\R}{\mathbb{R}}
\newcommand{\Z}{\mathbb{Z}}
\newcommand{\N}{\mathbb{N}}

\newcommand{\conv}[1]{\operatorname{conv}\left(#1\right)}

\newcommand{\sP}{\mathsf{P}}
\newcommand{\sQ}{\mathsf{Q}}
\newcommand{\sON}{\mathsf{ON}}
\newcommand{\sOFF}{\mathsf{OFF}}
\newcommand{\bmu}{\bm{\mu}}
\newcommand{\bbeta}{\bm{\eta}}
\newcommand{\bomega}{\bm{\omega}}
\newcommand{\bphi}{\bm{\phi}}
\newcommand{\blambda}{\bm{\lambda}}
\newcommand{\bv}{\bm{v}}

\newcommand{\bU}{\bm{U}}
\newcommand{\cvec}[1]{\begin{bmatrix} #1 \end{bmatrix}}
\newcommand{\cN}{\mathcal{N}}
\newcommand{\1}[1]{1_{[#1]}}
\newcommand{\Mone}{\mathsf{M1}}
\newcommand{\Mtwo}{\mathsf{M2}}
\newcommand{\hI}{\hat{I}}
\newcommand{\hi}{\hat{i}}

\begin{document}

\title{Exploiting Channel Memory for Multi-User Wireless Scheduling without Channel Measurement: Capacity Regions and Algorithms}



\author{\large{Chih-ping~Li,~\IEEEmembership{Student Member,~IEEE} and Michael~J.~Neely,~\IEEEmembership{Senior Member,~IEEE}}%
\thanks{Chih-ping Li (web: http://www-scf.usc.edu/$\sim$chihpinl) and Michael J. Neely (web: http://www-rcf.usc.edu/$\sim$mjneely) are with the Department of Electrical Engineering, University of Southern California, Los Angeles, CA 90089, USA.}    %
\thanks{This material is supported in part  by one or more of the following: the DARPA IT-MANET program grant W911NF-07-0028, the NSF Career grant CCF-0747525, and continuing through participation in the Network Science Collaborative Technology Alliance sponsored by the U.S. Army Research Laboratory.}%
\thanks{This paper appears in part in~\cite{LaN10conf-channelmemory}.}
}

\markboth{}{}
\maketitle

\begin{abstract}
We study the fundamental network capacity of a multi-user
wireless downlink under two assumptions: (1) Channels
are not explicitly measured and thus instantaneous states
are unknown, (2) Channels are modeled as $\sON/\sOFF$
Markov chains.  This is an important network model to explore because
channel probing may be costly or infeasible in some contexts.
In this case, we can use channel memory with ACK/NACK feedback
from previous transmissions to improve network throughput.
Computing in closed form the capacity region of this network
is difficult because it involves solving a high dimension partially
observed Markov decision problem. Instead, in this paper we
construct an inner and outer bound on the capacity region,
showing that the bound is tight when the number of users is
large and the traffic is symmetric. For the case of heterogeneous
traffic and any number of users, we propose a simple
queue-dependent policy that can stabilize the network with any
data rates strictly within the inner capacity bound. The stability
analysis uses a novel frame-based Lyapunov drift argument.
The outer-bound analysis uses stochastic coupling and
state aggregation to bound the performance of a restless
bandit problem using a related multi-armed bandit system.
Our results are useful in cognitive radio
networks, opportunistic scheduling with delayed/uncertain
channel state information, and restless bandit problems.
\end{abstract}


\section{Introduction} \label{sec:intro}
\IEEEPARstart{D}{ue} to the increasing demand of cellular network services,  in the past fifteen years efficient communication over a single-hop wireless downlink has been extensively studied. In this paper we study the fundamental network capacity of a time-slotted wireless downlink under the following assumptions: (1) Channels are never explicitly probed, and thus their instantaneous  states are never known, (2) Channels are modeled as two-state $\sON$/$\sOFF$ Markov chains.   This network model is  important because, due to the energy and timing overhead, learning instantaneous channel states by probing may  be costly or infeasible. Even if this is feasible (when channel coherence time is relatively large), the time consumed by channel probing cannot be re-used for data transmission, and transmitting data without probing  may achieve higher throughput~\cite{LaN10}.
\footnote{
One quick example is to consider a time-slotted channel with state space $\{\mathsf{B}, \mathsf{G}\}$. Suppose channel states are i.i.d. over slots with stationary probabilities $\prob{\mathsf{B}} = 0.2$ and $\prob{\mathsf{G}}=0.8$. At state $\mathsf{B}$ and $\mathsf{G}$, at most $1$ and $2$ packets can be successfully delivered in a slot, respectively. Packet transmissions beyond the capacity will all fail and need retransmissions. Channel probing can be done on each slot, which consumes $0.2$ fraction of a slot. Then the policy that always probes the channel yields throughput $0.8(2\cdot 0.8 + 1\cdot 0.2) = 1.44$, while the policy that never probes the channel and always sends packets at rate $2$ packets/slot yields throughput $2\cdot 0.8 = 1.6 > 1.44$.
}
In addition, since wireless channels can be adequately modeled as Markov chains~\cite{WaC96,  ZRM96conf}, we shall take advantage of channel memory to improve network throughput.

Specifically, we consider a time-slotted wireless downlink where a base station serves $N$ users through $N$ (possibly different) \emph{positively correlated} Markov $\sON$/$\sOFF$ channels. 
Channels are never probed so that their instantaneous states are unknown. In every slot, the base station selects at most one user to which it transmits a packet. We assume every packet transmission takes exactly one slot. Whether the transmission succeeds depends on the unknown state of the channel. At the end of a slot, an ACK/NACK is fed back from the served user to the base station. Since channels are either $\sON$ or $\sOFF$, this feedback reveals the channel state of the served user in the last slot and provides partial information of future states. Our goal is to characterize all achievable throughput vectors in this network, and to design simple throughput-achieving algorithms.

We define the \emph{network capacity region} $\Lambda$ as the closure of the set of all achievable throughput vectors. We can compute $\Lambda$ by locating its boundary points. Every boundary point can be computed by formulating a partially observable Markov decision process (POMDP)~\cite{Ber05book}, with information states defined as, conditioning on the channel observation history,  the probabilities that channels are $\sON$. This approach, however, is computationally prohibitive because the information state space is countably infinite (which we will show later) and grows exponentially fast with $N$.

The first contribution of this paper is that we construct an outer and an inner bound on  $\Lambda$. The outer bound comes from analyzing a fictitious channel model in which every scheduling policy yields higher throughput than it does in the real network.  The inner bound is the achievable rate region of a special class of \emph{randomized round robin policies} (introduced in Section~\ref{sec:401}). These policies are simple and take advantage of channel memory.  In the case of symmetric channels (that is, channels are i.i.d.) and when the network serves a large number of users, we show that as data rates are more \emph{balanced}, or in a geometric sense as the direction of the data rate vector in the Euclidean space is closer to the $45$-degree angle, the inner bound converges geometrically fast to the outer bound, and the bounds are tight.  This analysis uses results in~\cite{ZKL08,ALJ09} that derive an outer bound on the maximum sum throughput for a symmetric system.

The inner capacity bound is indeed useful. ÊFirst, the structure of the bound itself shows how channel memory improves throughput. ÊSecond, we show analytically that a large class of intuitively good heuristic policies achieve throughput that is at least as good as this bound, and hence the bound acts as a (non-trivial) performance guarantee. Finally, supporting throughput outside this bound may inevitably involve solving a much more complicated POMDP. ÊThus,
for simplicity and practicality, we may regard the inner bound as an \emph{operational} network capacity region.

In this paper we also derive a simple queue-dependent dynamic round robin policy that stabilizes the network whenever the arrival rate vector is interior to our inner bound. This policy has polynomial time complexity and is derived by a novel \emph{variable-length frame-based Lyapunov analysis}, first used in~\cite{Nee09conf} in a different
context. This analysis is important because the inner bound is based on a mixture of many different types of round robin policies, and an offline computation of the
proper time average mixtures needed to achieve a given point
in this complex inner bound would require solving $\Theta(2^N)$ unknowns in a linear system, which is impractical when $N$ is large. ÊThe Lyapunov analysis overcomes this complexity difficulty with online queue-dependent decisions.

The results of this paper apply to the emerging area of  opportunistic spectrum access in cognitive radio networks (see~\cite{ZaS07_2} and references therein), where the channel occupancy of a primary user acts as a Markov $\sON$/$\sOFF$ channel to the secondary users. Specifically, our results apply to the important case where each of the secondary users has a designated channel and they cooperate via a centralized controller. This paper is also a study on efficient scheduling over wireless networks with delayed/uncertain channel state information (CSI) (see~\cite{PET07conf, YaS08conf, YaS09conf} and references therein). The work on delayed CSI that is most closely related to ours is~\cite{YaS08conf, YaS09conf}, where the authors study the capacity region and throughput-optimal policies of different wireless networks, assuming that channel states are \emph{persistently} probed but fed back \emph{with delay}. We note that our paper is significantly different. Here channels are never probed, and new (delayed) CSI of a channel is only acquired when the channel is served. Implicitly, acquiring the delayed CSI of any channel is part of the control decisions in this paper.

This paper is organized as follows. The network model is given in Section~\ref{sec:model}, inner and outer bounds are constructed in Sections~\ref{sec:502} and~\ref{sec:503}, and compared in Section~\ref{sec:801} in the case of symmetric channels. Section~\ref{sec:501} gives the queue-dependent policy to achieve the inner bound.

\section{Network Model} \label{sec:model}
\begin{figure}[htbp]
\centering
\includegraphics[width=3in]{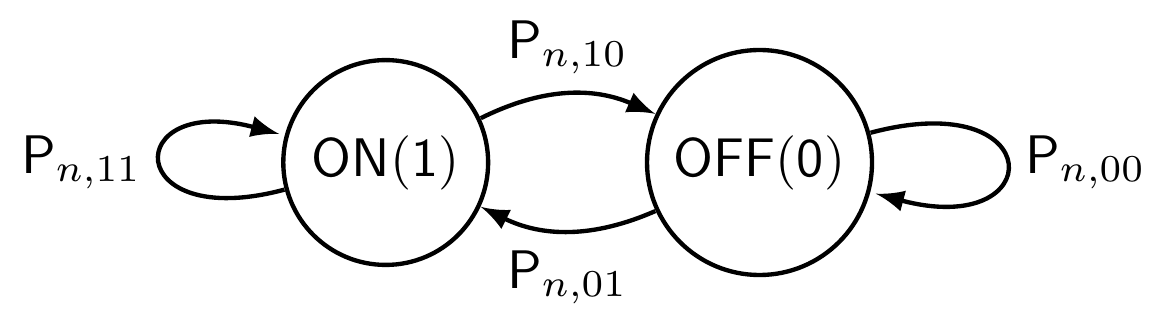}
\caption{A two-state Markov $\sON$/$\sOFF$ chain for channel $n\in\{1, 2, \ldots, N\}$.}
\label{fig:markovonoffchain}
\end{figure}

Consider a base station transmitting data to $N$ users through $N$ Markov $\sON$/$\sOFF$ channels. Suppose time is slotted with normalized slots $t$ in $\{0, 1, 2, \ldots\}$. Each channel is modeled as a two-state $\sON$/$\sOFF$ Markov chain (see Fig.~\ref{fig:markovonoffchain}).
The state evolution of channel $n\in\{1, 2, \ldots, N\}$  follows the transition probability matrix
\[
\bm{\sP}_{n}
=
\begin{bmatrix}
\sP_{n, 00} & \sP_{n, 01} \\
\sP_{n, 10} & \sP_{n, 11}
\end{bmatrix},
\]
where state $\sON$ is represented by $1$ and $\sOFF$ by $0$, and $\sP_{n, ij}$ denotes the transition probability from state $i$ to $j$. We assume $\sP_{n, 11}<1$ for all $n$ so that no channel is constantly $\sON$. Incorporating constantly $\sON$ channels like wired links is easy and thus omitted in this paper. We suppose channel states are fixed in every slot and may only change at slot boundaries. We assume all channels are positively correlated, which, in terms of transition probabilities, is equivalent to assuming $\sP_{n, 11} > \sP_{n, 01}$ or $\sP_{n, 01}+ \sP_{n, 10} < 1$ for all $n$.\footnote{Assumption $\sP_{n, 11} > \sP_{n, 01}$ yields that the state $s_{n}(t)$ of channel $n$  has auto-covariance  $\expect{(s_{n}(t) - \mathbb{E}s_{n}(t))(s_{n}(t+1)-\mathbb{E}s_{n}(t+1))} >0$. In addition, we note that the case $\sP_{n,11} = \sP_{n,01}$ corresponds to a channel having i.i.d. states over slots. Although we can naturally incorporate i.i.d. channels into our model and all our results still hold, we exclude them in this paper because we shall show how throughput can be improved by channel memory, which i.i.d. channels do not have. The degenerate case where all channels are i.i.d. over slots is fully solved in~\cite{LaN10}.}  We suppose  the base station keeps $N$ queues of infinite capacity to store exogenous packet arrivals destined for the $N$ users. At the beginning of every slot, the base station attempts to transmit a packet (if there is any) to a selected user. We suppose the base station has no channel probing capability and must select users oblivious of the current channel states. If a user is selected and its current channel state is $\sON$, one packet is successfully delivered to that user. Otherwise, the transmission fails and zero packets are served. At the end of a slot in which the base station serves a user,  an ACK/NACK message is fed back from the selected user to the base station through an independent error-free control channel, according to whether the transmission succeeds. Failing to receive an ACK is regarded as a NACK. Since channel states are either $\sON$ or $\sOFF$, such feedback reveals the channel state of the selected user in the last slot.

Conditioning on all past channel observations, define the $N$-dimensional \emph{information state vector} $\bomega(t) = (\omega_{n}(t): 1\leq n\leq N)$ where $\omega_{n}(t)$ is the conditional probability  that channel $n$ is $\sON$ in slot $t$. We assume initially $\omega_{n}(0) = \pi_{n, \sON}$ for all $n$, where $\pi_{n, \sON}$ denotes the stationary probability that channel $n$ is $\sON$. As discussed in~\cite[Chapter $5.4$]{Ber05book}, vector $\bomega(t)$ is a \emph{sufficient statistic}.  That is, instead of tracking the whole system history, the base station can act optimally only based on $\bomega(t)$. The base station shall keep track of the $\{\bomega(t)\}$ process.

We assume transition probability matrices $\bm{\sP}_{n}$ for all $n$ are known to the base station. We denote by $s_{n}(t)\in\{\sOFF, \sON\}$ the state of channel $n$ in slot $t$. Let $n(t)\in\{1, 2, \ldots, N\}$ denote the user  served in slot $t$. Based on the ACK/NACK feedback, vector $\bomega(t)$ is updated as follows. For $1\leq n\leq N$,
\begin{equation} \label{eq:801}
\omega_{n}(t+1) = 
\begin{cases}
\sP_{n, 01},\quad \text{if $n=n(t)$, $s_{n}(t)=\sOFF$} \\
\sP_{n, 11},\quad \text{if $n=n(t)$, $s_{n}(t)=\sON$} \\
\omega_{n}(t) \sP_{n, 11} + (1-\omega_{n}(t))\sP_{n, 01},\;\text{if $n\neq n(t)$}.
\end{cases}
\end{equation}
If in the most recent use of channel $n$, we observed (through feedback) its state was $i\in\{0,1\}$ in slot $(t-k)$ for some $k \leq t$, then $\omega_{n}(t)$ is equal to the $k$-step transition probability $\sP_{n, i1}^{(k)}$. In general, for any fixed $n$, probabilities $\omega_{n}(t)$ take values in the countably infinite set $\mathcal{W}_{n} = \{\sP_{n, 01}^{(k)}, \sP_{n, 11}^{(k)} : k\in\N\} \cup \{\pi_{n, \sON}\}$.  By eigenvalue decomposition on $\bm{\sP}_{n}$~\cite[Chapter $4$]{Gal96book}, we can show the $k$-step transition probability matrix $\bm{\sP}_{n}^{(k)}$ is
\begin{equation} \label{eq:201}
\begin{split}
&\bm{\sP}_{n}^{(k)} \triangleq
\begin{bmatrix}
\sP_{n, 00}^{(k)} & \sP_{n, 01}^{(k)} \\
\sP_{n, 10}^{(k)} & \sP_{n, 11}^{(k)}
\end{bmatrix}  
= \left(\bm{\sP}_{n}\right)^{k} \\
&=
\frac{1}{x_{n}} \begin{bmatrix} \mathsf{P}_{n, 10} + \mathsf{P}_{n, 01} (1-x_{n})^{k} & \mathsf{P}_{n, 01}\,(1-(1-x_{n})^{k}) \\ \mathsf{P}_{n, 10} (1-(1-x_{n})^{k}) & \mathsf{P}_{n, 01} + \mathsf{P}_{n, 10} (1-x_{n})^{k}\end{bmatrix},
\end{split}
\end{equation}
where we have defined $x_{n}\triangleq \sP_{n, 01}+\sP_{n, 10}$. Assuming that  channels are positively correlated, i.e., $x_{n}<1$, by~\eqref{eq:201} we have the following lemma.
\begin{lem} \label{lem:101}
For a positively correlated $(\sP_{n, 11}>\sP_{n, 01})$ Markov $\sON$/$\sOFF$ channel with transition probability matrix $\bm{\sP}_{n}$, we have
\begin{enumerate}
\item The stationary probability $\pi_{n, \sON} = \sP_{n, 01}/ x_{n}$.
\item The $k$-step transition probability $\sP_{n, 01}^{(k)}$ is nondecreasing in $k$ and $\sP_{n, 11}^{(k)}$ nonincreasing in $k$.  Both $\sP_{n, 01}^{(k)}$ and $\sP_{n, 11}^{(k)}$ converge to $\pi_{n, \sON}$ as $k\to\infty$.
\end{enumerate} 
\end{lem}

\begin{figure}[htbp]
\centering
\includegraphics[width=2in]{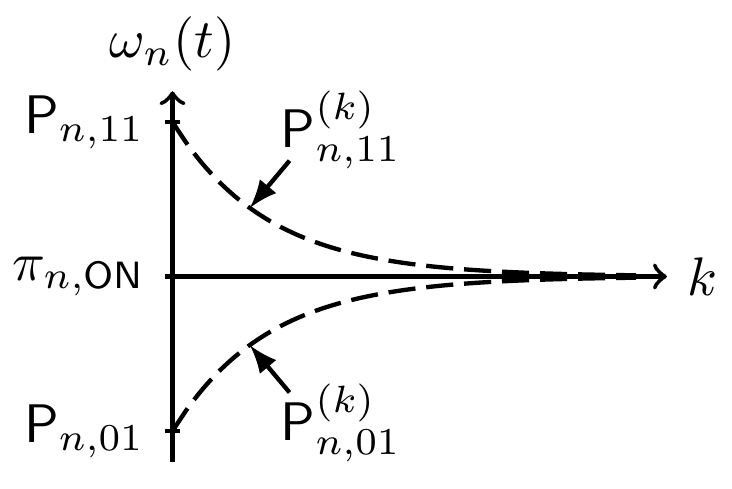}
\caption{Diagram of the $k$-step transition probabilities $\sP_{n,01}^{(k)}$ and $\sP_{n,11}^{(k)}$ of a positively correlated Markov $\sON/\sOFF$ channel.}
\label{fig:j102}
\end{figure}

As a corollary of Lemma~\ref{lem:101}, it follows that
\begin{equation} \label{eq:202}
\sP_{n, 11} \geq \sP_{n, 11}^{(k_{1})} \geq \sP_{n, 11}^{(k_{2})} \geq \pi_{n, \sON} \geq \sP_{n, 01}^{(k_{3})}  \geq \sP_{n, 01}^{(k_{4})} \geq \sP_{n, 01}
\end{equation}
for any integers $k_{1} \leq k_{2}$ and $k_{3} \geq k_{4}$ (see Fig.~\ref{fig:j102}). To maximize network throughput,~\eqref{eq:202} has some fundamental implications. We note that  $\omega_{n}(t)$ represents the transmission success probability over channel $n$ in slot $t$. Thus we shall keep serving a channel whenever its information state is $\sP_{n, 11}$, for it is the best state possible. Second, given that a channel was $\sOFF$ in its last use,  its information state improves as long as  the channel remains idle. Thus we shall wait as long as possible before reusing such a channel. Actually, when channels are symmetric ($\bm{\sP}_{n} = \bm{\sP}$ for all $n$), it is shown that a myopic policy  with this structure maximizes the sum throughput of the network~\cite{ALJ09}.

\section{A Round Robin Policy} \label{sec:502}
For any integer $M\in\{1, 2, \ldots, N\}$, we present a special round robin policy $\mathsf{RR}(M)$ serving the first $M$ users $\{1, 2, \ldots, M\}$ in the network. The $M$ users are served in the circular order $1 \!\to\! 2\!\to\! \cdots  \!\to\! M \!\to\! 1 \!\to\! \cdots$. In general, we can use this policy to serve any subset of users. This policy is the fundamental building block of all the results in this paper.

\subsection{The Policy}

\underline{\bf Round Robin Policy $\mathsf{RR}(M)$ }:

\begin{enumerate}
\item At time $0$, the base station starts with channel $1$. Suppose initially $\omega_{n}(0) = \pi_{n,\sON}$ for all $n$.
\item Suppose at time $t$, the base station switches to channel $n$. Transmit a \emph{data} packet to user $n$ with probability $\sP_{n,01}^{(M)} / \omega_{n}(t)$ and a \emph{dummy} packet otherwise. In both cases, we receive ACK/NACK information at the end of the slot. \label{item:2}
\item At time $(t+1)$, if a dummy packet is sent at time $t$, switch to channel $(n\mod M)+1$ and go to Step~\ref{item:2}. Otherwise, keep transmitting data packets over channel $n$ until we receive a NACK. Then switch to channel $(n \mod M)+1$ and go to Step~\ref{item:2}. We note that dummy packets are only sent on the first slot every time the base station switches to a new channel. 

\item Update $\bomega(t)$ according to~\eqref{eq:801} in every slot.
\end{enumerate}

Step~\ref{item:2} of $\mathsf{RR}(M)$ only makes sense if $\omega_{n}(t) \geq \sP_{n,01}^{(M)}$, which we prove in the next lemma.
\begin{lem} \label{lem:201}
Under $\mathsf{RR}(M)$, whenever the base station switches to channel $n\in\{1, 2, \ldots, M\}$ for another round of transmission, its current information state satisfies $\omega_{n}(t) \geq \sP_{n,01}^{(M)}$.
\end{lem}

\begin{IEEEproof}[Proof of Lemma~\ref{lem:201}]
See Appendix~\ref{pf:105}.
\end{IEEEproof}

We note that policy $\mathsf{RR}(M)$ is very conservative and not throughput-optimal. For example, we can improve the throughput by always sending data packets but no dummy ones. Also, it does not follow the guidelines we provide at the end of Section~\ref{sec:model} for maximum throughput. Yet, we will see that, in the case of symmetric channels, throughput under $\mathsf{RR}(M)$ is close to optimal when $M$ is large. Moreover, the underlying analysis of $\mathsf{RR}(M)$ is tractable so that we can mix such round robin policies over different subsets of users to form a non-trivial inner capacity bound. The tractability of $\mathsf{RR}(M)$ is because  it is equivalent to the following \emph{fictitious} round robin policy (which can be proved as a corollary of Lemma~\ref{lem:202} provided later).

\underline{\bf Equivalent Fictitious Round Robin}:
\begin{enumerate}
\item At time $0$, start with channel $1$.
\item When the base station switches to channel $n$, \emph{set its current information state to $\sP_{n,01}^{(M)}$}.\footnote{In reality we cannot \emph{set} the information state of a channel, and therefore the policy is fictitious.} Keep transmitting data packets over channel $n$ until we receive a NACK. Then switch to channel $(n \mod M)+1$ and repeat Step~\ref{item:3}. \label{item:3}
\end{enumerate}

For any round robin policy that serves channels in the circular order $1 \!\to\! 2\!\to\! \cdots  \!\to\! M \!\to\! 1 \!\to\! \cdots$, the technique of resetting the information state to $\sP_{n,01}^{(M)}$ creates a system with an information state that is \emph{worse} than the information state under the actual system. To see this, since in the actual system channels are served in the circular order, after we switch away from serving a particular channel $n$, we serve the other $(M-1)$ channels for at least one slot each, and so we return to channel $n$ after at least $M$ slots. Thus, its starting information state is always at least $\sP_{n,01}^{(M)}$ (the proof is similar to that of Lemma~\ref{lem:201}). Intuitively, since information states represent the packet transmission success probabilities, resetting them to lower values degrades throughput. This is the reason why our inner capacity bound constructed later  using $\mathsf{RR}(M)$ provides a throughput lower bound for a large class of policies.


\subsection{Network Throughput under $\mathsf{RR}(M)$} \label{sec:rrm-throughput}
Next we analyze the throughput vector achieved by $\mathsf{RR}(M)$.

\subsubsection{General Case}

Under $\mathsf{RR}(M)$, let $L_{kn}$ denote the duration of the $k$th time the base station stays with channel $n$. A sample path of the $\{L_{kn}\}$ process is
\begin{equation} \label{eq:j108}
(
\underbrace{L_{11}, L_{12}, \ldots, L_{1M}}_{\text{round $k=1$}},
\underbrace{L_{21}, L_{22}, \ldots, L_{2M}}_{\text{round $k=2$}},
L_{31}, \ldots).
\end{equation}
The next lemma presents useful properties of $L_{kn}$, which serve as the foundation of the throughput analysis in the rest of the paper.
\begin{lem} \label{lem:202}
For any integer $k$ and $n\in\{1, 2, \ldots, M\}$,
\begin{enumerate}
\item The probability mass function of $L_{kn}$ is independent of $k$, and is
\[
L_{kn} = \begin{cases}
1 & \text{with prob. } 1-\sP_{n,01}^{(M)} \\
j\geq 2 & \text{with prob. }  \sP_{n,01}^{(M)}\,(\sP_{n,11})^{(j-2)}\,\sP_{n,10}.
\end{cases}
\]
As a result, for all $k\in\N$ we have
\[
\expect{L_{kn}} = 1+\frac{\sP_{n,01}^{(M)}}{\sP_{n,10}} = 
1+ \frac{
	\sP_{n,01} (1 - (1-x_{n}))^{M}
}{
	x_{n} \sP_{n,10}
}.
\]
\item  The number of data packets served in $L_{kn}$ is $(L_{kn}-1)$.
\item For every fixed channel $n$,  time durations $L_{kn}$ are \emph{i.i.d.} random variables over all $k$. 
\end{enumerate}
\end{lem}
\begin{IEEEproof}[Proof of Lemma~\ref{lem:202}]
\begin{enumerate}
\item Note that $L_{kn} = 1$ if, on the first slot of serving channel $n$, either a dummy packet is transmitted or a data packet is transmitted but the channel is $\sOFF$. This event occurs with probability
\[
\left(
1-\frac{
	\sP_{n,01}^{(M)}
}{
	\omega_{n}(t)
}
\right)
+ 
\frac{\sP_{n,01}^{(M)}}{\omega_{n}(t)} \left( 1-\omega_{n}(t)\right)
= 1 - \sP_{n,01}^{(M)}.
\]
Next, $L_{kn} = j \geq 2$ if in the first slot a data packet is successfully served, and this is followed by $(j-2)$ consecutive $\sON$ slots and one $\sOFF$ slot. This happens with probability  $\sP_{n,01}^{(M)}\,(\sP_{n,11})^{(j-2)}\,\sP_{n,10}$. The expectation of $L_{kn}$ can be directly computed from the probability mass function.

\item We can observe that one data packet is served in every slot of $L_{kn}$ except for the last one (when a dummy packet is sent over channel $n$, we have $L_{kn}=1$ and zero data packets are served).

\item At the beginning of every $L_{kn}$, we observe from the equivalent fictitious round robin policy that $\mathsf{RR}(M)$ effectively fixes $\sP_{n,01}^{(M)}$ as the current information state, regardless of the true current state $\omega_{n}(t)$. Neglecting $\omega_{n}(t)$ is to discard all system history, including all past $L_{k'n}$ for all $k' < k$. Thus $L_{kn}$ are i.i.d.. Specifically, for any $k'<k$  and integers~$l_{k'}$ and~$l_{k}$ we have
\[
\prob{L_{kn} = l_{k} \mid L_{k'n} = l_{k'}} = \prob{L_{kn} = l_{k}}.
\]
\end{enumerate}
\end{IEEEproof}

Now we can derive the throughput vector supported by $\mathsf{RR}(M)$. Fix an integer $K>0$. By Lemma~\ref{lem:202},  the time average throughput over channel $n$ after all channels finish their $K$th rounds, which we denote by $\mu_{n}(K)$, is
\[
\mu_{n}(K) \triangleq
\frac{
	\sum_{k=1}^{K} (L_{kn} - 1)
}{
	\sum_{k=1}^{K} \sum_{n=1}^{M} L_{kn}
}.
\]
Passing $K\to\infty$, we get
\begin{equation} \label{eq:204}
\begin{split}
&\lim_{K\to\infty} \mu_{n}(K) \\
&=
\lim_{K\to\infty}
\frac{
	\sum_{k=1}^{K} (L_{kn} - 1)
}{
	\sum_{k=1}^{K} \sum_{n=1}^{M} L_{kn}
} \\
&=
\lim_{K\to\infty}
\frac{
	(1/K) \sum_{k=1}^{K}  \left(L_{kn} - 1\right)
}{
	\sum_{n=1}^{M}  (1/K) \sum_{k=1}^{K}  L_{kn}
} \\
&\stackrel{(a)}{=}
\frac{\expect{L_{1n}}-1}{\sum_{n=1}^{M} \expect{L_{1n}}} \\
&\stackrel{(b)}{=}
\frac{
	\sP_{n,01} (1-(1-x_{n})^{M}) / (x_{n} \sP_{n,10})
}{
	M + \sum_{n=1}^{M} \sP_{n,01} (1-(1-x_{n})^{M}) / (x_{n} \sP_{n,10})
},
\end{split}
\end{equation}
where $(a)$ is by the Law of Large Numbers (noting by Lemma~\ref{lem:202} that $L_{kn}$ are i.i.d. over $k$), and $(b)$ is by Lemma~\ref{lem:202}.

\subsubsection{Symmetric Case}

We are particularly interested in the sum throughput under $\mathsf{RR}(M)$ when channels are symmetric,
that is, all channels have the same statistics $\bm{\sP}_{n} = \bm{\sP}$ for all $n$. In this case, by channel symmetry every channel has the same throughput. From~\eqref{eq:204}, we can show the sum throughput is
\[
\sum_{n=1}^{M} \lim_{K\to\infty} \mu_{n}(K) = \frac{
	\sP_{01}(1-(1-x)^{M})
}{
	x\,\sP_{10} + \sP_{01}(1-(1-x)^{M})
},
\]
where in the last term the subscript $n$ is dropped due to channel symmetry. It is handy to define a function $c_{(\cdot)}: \N \to \R$ as
\begin{equation} \label{eq:j101}
c_{M} \triangleq \frac{
	\sP_{01}(1-(1-x)^{M})
}{
	x\,\sP_{10} + \sP_{01}(1-(1-x)^{M})
}, \quad x \triangleq \sP_{01}+\sP_{10},
\end{equation}
and define $c_{\infty} \triangleq \lim_{M\to\infty} c_{M} = \sP_{01}/(x\sP_{10}+\sP_{01})$ (note that $x<1$ because every channel is positively correlated over time slots). The function $c_{(\cdot)}$ will be used extensively in this paper. We summarize the above derivation in the next lemma.
\begin{lem} \label{lem:203}
Policy $\mathsf{RR}(M)$ serves channel $n\in\{1, 2, \ldots, M\}$ with throughput
\[
\frac{
	\sP_{n,01} (1-(1-x_{n})^{M}) / (x_{n} \sP_{n,10})
}{
	M + \sum_{n=1}^{M} \sP_{n,01} (1-(1-x_{n})^{M}) / (x_{n} \sP_{n,10})
}.
\]
In particular, in symmetric channels the sum throughput under $\mathsf{RR}(M)$ is $c_{M}$ defined as
\[
c_{M} = \frac{
	\sP_{01}(1-(1-x)^{M})
}{
	x\,\sP_{10} + \sP_{01}(1-(1-x)^{M})
}, \quad x = \sP_{01} + \sP_{10},
\]
and every channel has throughput $c_{M}/M$.
\end{lem}

We remark that the sum throughput $c_{M}$ of $\mathsf{RR}(M)$ in the symmetric case is nondecreasing in $M$, and thus can be improved by serving more channels. Interestingly, here we see that the sum throughput is improved by having \emph{multiuser diversity} in the network, even though  instantaneous channel states are never known.

\subsection{How Good is $\mathsf{RR}(M)$?} \label{sec:pointproximity}
Next, in symmetric channels, we quantify how close the sum throughput $c_{M}$  is to optimal. The following lemma presents a useful upper bound on the maximum sum throughput.

\begin{lem}[\cite{ZKL08,ALJ09}] \label{lem:108}
In symmetric channels, any scheduling policy that confines to our model has sum throughput less than or equal to $c_{\infty}$.\footnote{We note that the throughput analysis in~\cite{ZKL08} makes a minor assumption on the existence of some limiting time average.  Using similar ideas of~\cite{ZKL08}, in Theorem~\ref{thm:401} of Section~\ref{sec:generalouterbound} we will construct an upper bound on the maximum sum throughput for general positively correlated Markov $\sON$/$\sOFF$ channels. When restricted to the symmetric case, we get the same upper bound without any assumption.}
\end{lem}

By Lemma~\ref{lem:203} and~\ref{lem:108}, the loss of the sum throughput of $\mathsf{RR}(M)$ is no larger than  $c_{\infty} - c_{M}$. Define $\widetilde{c}_{M}$ as
\[
\widetilde{c}_{M}
\triangleq 
\frac{\sP_{01}(1-(1-x)^{M})}{x \sP_{10}+\sP_{01}}
= c_{\infty} (1-(1-x)^{M})
\]
and note that $\widetilde{c}_{M} \leq c_{M} \leq c_{\infty}$. It follows
\begin{equation} \label{eq:222}
c_{\infty} - c_{M} \leq c_{\infty} - \widetilde{c}_{M} = c_{\infty}(1-x)^{M}.
\end{equation}
The last term of~\eqref{eq:222} decreases to zero geometrically fast as $M$ increases. This indicates that  $\mathsf{RR}(M)$ yields near-optimal sum throughput  even  when it only serves a moderately large number of channels.

\section{Randomized Round Robin Policy, Inner and Outer Capacity Bound} \label{sec:503}

\subsection{Randomized Round Robin Policy} \label{sec:401}
%
Lemma~\ref{lem:203} specifies the throughput vector achieved by implementing
$\mathsf{RR}(M)$ over a particular collection of $M$ channels.  Here we are
interested in the set of throughput vectors achievable by randomly
mixing $\mathsf{RR}(M)$-like policies over different channel subsets and allowing
a different round-robin ordering on each subset.  To generalize
the $\mathsf{RR}(M)$ policy, first let $\Phi$ denote the set of all $N$-dimensional binary vectors excluding the all-zero vector $(0, 0, \ldots, 0)$. For any binary vector $\bphi = (\phi_{1}, \phi_{2}, \ldots, \phi_{N})$ in $\Phi$, we say channel $n$ is \emph{active} in $\bphi$ if $\phi_{n} = 1$. Each vector $\bphi\in\Phi$ represents a different subset of active channels. We denote by $M(\bphi)$ the number of active channels in $\bphi$. 

For each $\bphi\in\Phi$, consider the following round robin policy $\mathsf{RR}(\bphi)$ that serves active channels in $\bphi$ in every round.

\underline{\bf Dynamic Round Robin Policy $\mathsf{RR}(\bphi)$}:
\begin{enumerate}
\item \emph{Deciding the service order in each round}: 

At the beginning of each round, we denote by $\tau_{n}$ the time duration between the last use of channel $n$ and the beginning of the current round. Active channels in $\bphi$ are served in the decreasing order of $\tau_{n}$ in this round (in other words, the active channel that is \emph{least recently used} is served first). \label{item:6}

\item \emph{On each active channel in a round}: 
	\begin{enumerate}
	\item Suppose at time $t$ the base station switches to channel $n$. Transmit a \emph{data} packet to user $n$ with probability $\sP_{n,01}^{(M(\bphi))} / \omega_{n}(t)$ and a \emph{dummy} packet otherwise. In both cases, we receive ACK/NACK information at the end of the slot. \label{item:1}
 	\item At time $(t+1)$, if a dummy packet is sent at time $t$, switch to the next active channel following the order given in Step~\ref{item:6}. Otherwise, keep transmitting data packets over channel $n$ until we receive a NACK. Then switch to the next active channel and go to Step~\ref{item:1}. We note that dummy packets are only sent on the first slot every time the base station switches to a new channel.
	\end{enumerate}
\item Update $\bomega(t)$ according to~\eqref{eq:801} in every slot.
\end{enumerate}

Using $\mathsf{RR}(\bphi)$ as building blocks, we consider the following class of \emph{randomized round robin policies}.

\underline{{\bf Randomized Round Robin Policy} $\mathsf{RandRR}$}:
\begin{enumerate}
\item Pick $\bphi \in \Phi$ with probability $\alpha_{\bphi}$, where $\sum_{\bphi\in\Phi} \alpha_{\bphi} = 1$. \label{item:4}
\item Run policy $\mathsf{RR}(\bphi)$ for one round. Then go to Step~\ref{item:4}. \label{item:7}
\end{enumerate}


Note that active channels may be served in different order in different rounds, according to the least-recently-used service order. This allows more time for $\mathsf{OFF}$ channels to return to better information states (note that $\sP_{n,01}^{(k)}$ is nondecreasing in $k$) and thus improves throughput. The next lemma guarantees the feasibility of executing any $\mathsf{RR}(\bphi)$ policy in $\mathsf{RandRR}$ (similar to Lemma~\ref{lem:201}, whenever the base station switches to a new channel $n$, we need $\omega_{n}(t) \geq \sP_{n,01}^{(M(\bphi))}$ in Step~\ref{item:1} of $\mathsf{RR}(\bphi)$).

\begin{lem} \label{lem:106}
When $\mathsf{RR}(\bphi)$ is chosen by $\mathsf{RandRR}$ for a new round of transmission, every active channel $n$ in $\bphi$ starts with information state no worse than  $\sP_{n,01}^{(M(\bphi))}$.
\end{lem}

\begin{IEEEproof}[Proof of Lemma~\ref{lem:106}]
See Appendix~\ref{pf:103}.
\end{IEEEproof}

Although $\mathsf{RandRR}$ randomly selects subsets of users and serves them
in an order that depends on previous choices, we can surprisingly analyze
its throughput.  This is done by using the throughput analysis of $\mathsf{RR}(M)$, as shown
in the following corollary to Lemma~\ref{lem:202}:

\begin{cor} \label{cor:201}
For each policy $\mathsf{RR}(\bphi)$, $\bphi\in\Phi$, within time periods in which $\mathsf{RR}(\bphi)$ is executed by $\mathsf{RandRR}$, denote by $L_{kn}^{\bphi}$ the duration of the $k$th time  the base station stays with active channel $n$.  Then:
\begin{enumerate}
\item The probability mass function of $L_{kn}^{\bphi}$ is independent of $k$, and is
\[
L_{kn}^{\bphi} = \begin{cases}
1 & \!\text{with prob. } 1-\sP_{n,01}^{(M(\bphi))} \\
j\geq 2 & \!\text{with prob. }  \sP_{n,01}^{(M(\bphi))}\,(\sP_{n,11})^{(j-2)}\,\sP_{n,10}.
\end{cases}
\]
As a result, for all $k\in\N$ we have
\begin{equation} \label{eq:j105}
\begin{split}
\expect{L_{kn}^{\bphi}}  &= 1 + \frac{\sP_{n,01}^{(M(\bphi))}}{\sP_{n,10}}.
\end{split}
\end{equation}
\item  The number of data packets served in $L_{kn}^{\bphi}$ is $(L_{kn}^{\bphi}-1)$.
\item For every fixed $\bphi$ and every fixed active channel $n$ in $\bphi$, the time durations $L_{kn}^{\bphi}$ are i.i.d. random variables over all $k$.
\end{enumerate}
\end{cor}

\subsection{Achievable Network Capacity --- An Inner Capacity Bound} \label{sec:802}

Using Corollary~\ref{cor:201}, next we present the achievable rate region of the class of $\mathsf{RandRR}$ policies. For each $\mathsf{RR}(\bphi)$ policy, define an $N$-dimensional vector $\bbeta^{\bphi} = (\eta_{1}^{\bphi}, \eta_{2}^{\bphi}, \ldots, \eta_{N}^{\bphi})$ where
\begin{equation} \label{eq:509}
\eta_{n}^{\bphi} \triangleq \begin{cases} \frac{\expect{L_{1n}^{\bphi}}-1}{\sum_{n: \phi_{n} =1} \expect{L_{1n}^{\bphi}} } & \text{if channel $n$ is active in $\bphi$,} \\ 0 & \text{otherwise,} \end{cases}
\end{equation}
where $\expect{L_{1n}^{\bphi}}$ is given in~\eqref{eq:j105}.
Intuitively, by the analysis prior to Lemma~\ref{lem:203}, round robin policy $\mathsf{RR}(\bphi)$ yields throughput $\eta_{n}^{\bphi}$ over channel $n$ for each $n\in\{1, 2, \ldots, N\}$.
%
%
%
Incorporating all possible random mixtures of $\mathsf{RR}(\bphi)$ policies for different $\bphi$, $\mathsf{RandRR}$ can support any data rate vector that is entrywise dominated by a convex combination of vectors $\{\bbeta^{\bphi}\}_{\bphi\in\Phi}$ as shown by the next theorem. 

\begin{thm}[Generalized Inner Capacity Bound] \label{thm:205}
The class of $\mathsf{RandRR}$ policies supports all data rate vectors $\blambda$ in the set $\Lambda_{\text{int}}$ defined as
\[
\Lambda_{\text{int}} \triangleq \left\{
	\blambda \mid \bm{0}\leq \blambda \leq \bmu,\ \bmu\in\conv{
	\left\{ \bbeta^{\bphi}\right\}_{\bphi\in\Phi}
}
\right\},
\]
where  $\bbeta^{\bphi}$ is defined in~\eqref{eq:509}, $\conv{A}$ denotes the convex hull of set $A$, and $\leq$ is taken entrywise.
\end{thm}

\begin{IEEEproof}[Proof of Theorem~\ref{thm:205}]
See Appendix~\ref{pf:104}.
\end{IEEEproof}

Applying Theorem~\ref{thm:205} to symmetric channels yields the following corollary.

\begin{cor}[Inner Capacity Bound for Symmetric Channels] \label{cor:202}
In symmetric channels, the class of $\mathsf{RandRR}$ policies supports all rate vectors $\blambda\in\Lambda_{\text{int}}$ where
\[
\Lambda_{\text{int}} = \left\{
\blambda \mid
\bm{0}\leq \blambda \leq \bmu,\,\bmu\in\conv{ \left\{\frac{c_{M(\bphi)}}{M(\bphi)} \bphi \right\}_{\bphi\in\Phi}}
\right\},
\]
where $c_{M(\bphi)}$ is defined in~\eqref{eq:j101}.
\end{cor}

An example of the inner capacity bound and a simple queue-dependent dynamic policy that supports all data  rates within this nontrivial inner  bound will be provided later.

\subsection{Outer Capacity Bound} \label{sec:generalouterbound}

\begin{figure}[htbp]
\centering
\includegraphics[width=2.8in]{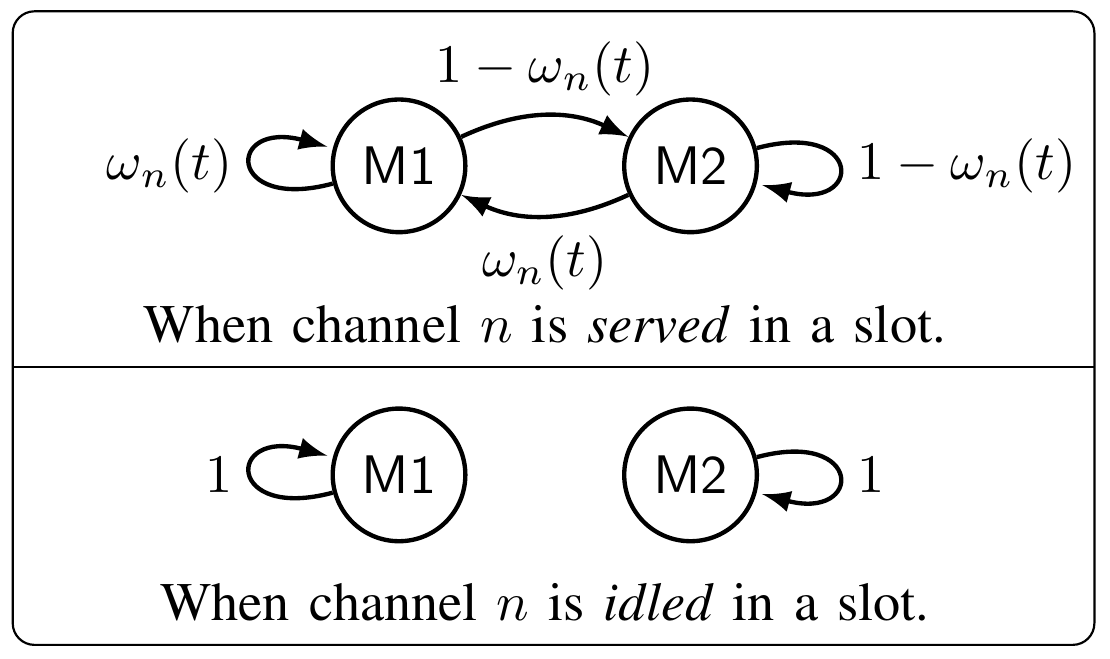}
\caption{Mode transition diagrams for the real channel $n$.}
\label{fig:501}
\end{figure}

\begin{figure}[htbp]
\centering
\includegraphics[width=2.8in]{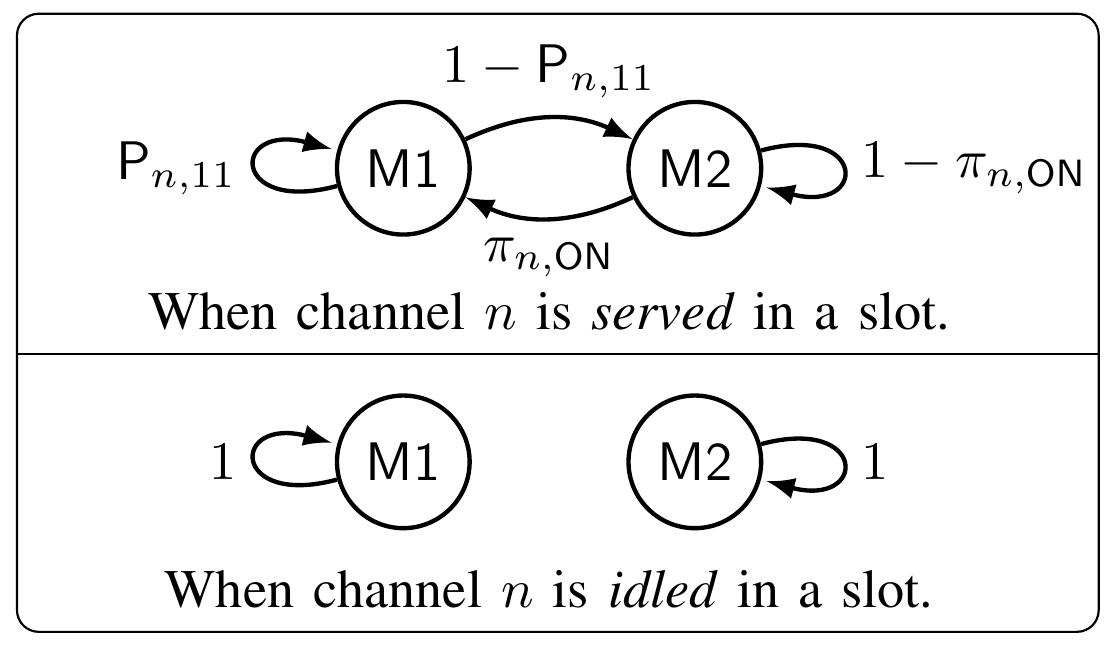}
\caption{Mode transition diagrams for the fictitious channel $n$.}
\label{fig:401}
\end{figure}

We construct an outer bound on $\Lambda$ using several novel ideas. First, by state aggregation, we transform the information state process $\{\omega_{n}(t)\}$ for each channel $n$ into non-stationary two-state Markov chains (in Fig.~\ref{fig:501} provided later).  Second, we create a set of bounding stationary Markov chains (in Fig.~\ref{fig:401} provided later),  which has the structure of a multi-armed bandit system. Finally, we create an outer capacity bound by relating the bounding model to the original non-stationary Markov chains using stochastic coupling. We note that since the control of the set of information state processes $\{\omega_{n}(t)\}$ for all $n$ can be viewed as a restless bandit problem~\cite{Whi88}, it is interesting to see how we bound the optimal performance of a restless bandit problem by a related multi-armed bandit system.

\begin{figure}[htbp]
\centering
\includegraphics[width=2.5in]{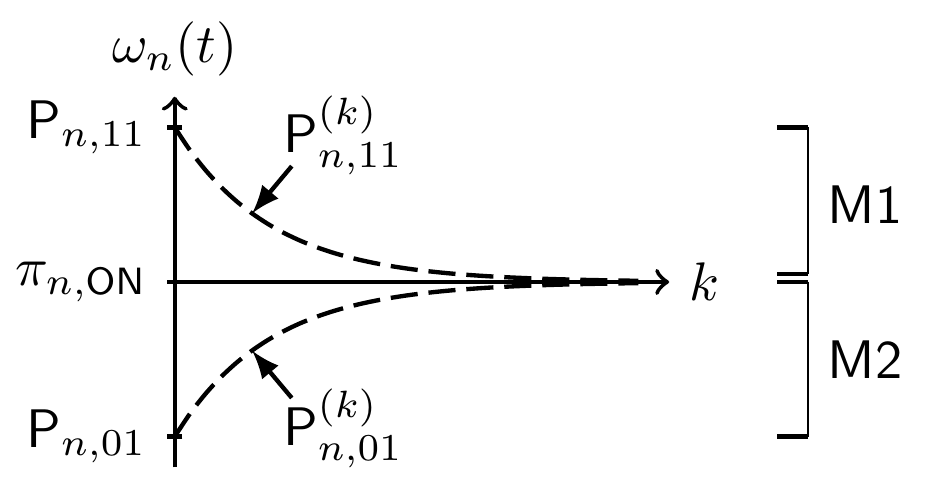}
\caption{The mapping $f_{n}$ from information states $\omega_{n}(t)$ to modes $\{\Mone, \Mtwo\}$.}
\label{fig:j501}
\end{figure}

We first \emph{map} channel information states  $\omega_{n}(t)$ into \emph{modes} for each $n\in\{1, 2, \ldots, N\}$. Inspired by~\eqref{eq:202}, we observe that each channel $n$ must be in one of the following two modes:
\begin{itemize}
\item[$\Mone$] The last observed state is $\sON$, and the channel has not been seen (through feedback) to turn $\sOFF$. In this mode the information state $\omega_{n}(t)\in[\pi_{n, \sON}, \sP_{n, 11}]$.
\item[$\Mtwo$] The last observed state is $\sOFF$, and the channel has not been seen to turned $\sON$. Here $\omega_{n}(t)\in[\sP_{n, 01}, \pi_{n, \sON}]$.
\end{itemize}
On channel $n$, recall that $\mathcal{W}_{n}$ is the state space of $\omega_{n}(t)$, and define a map $f_{n} : \mathcal{W}_{n} \to \{\Mone,\Mtwo\}$ where 
\[
f_{n}(\omega_{n}(t)) = \begin{cases} \Mone & \text{if $\omega_{n}(t) \in (\pi_{n,\sON}, \sP_{n,11}]$,} \\ \Mtwo & \text{if $\omega_{n}(t)\in[\sP_{n,01}, \pi_{n,\sON}]$.} \end{cases}
\]
This mapping is illustrated in Fig.~\ref{fig:j501}.

For any information state process $\{\omega_{n}(t)\}$ (controlled by some scheduling policy), the corresponding mode transition process under $f_{n}$ can be represented by the  Markov chains shown in Fig.~\ref{fig:501}. Specifically, when channel $n$ is served in a slot, the associated mode transition follows the upper non-stationary chain of Fig.~\ref{fig:501}. When channel $n$ is idled in a slot, the mode transition follows the lower stationary chain of Fig.~\ref{fig:501}. In the upper chain of Fig.~\ref{fig:501}, regardless what the current mode is, mode $\Mone$ is visited in the next slot if and only if channel $n$ is $\sON$ in the current slot, which occurs with probability $\omega_{n}(t)$. In the lower chain of Fig.~\ref{fig:501}, when channel $n$ is idled, its information state changes from a $k$-step transition probability to the $(k+1)$-step transition probability with the same most recent observed channel state. Therefore, the next mode stays the same as the current mode. We emphasize that, in the upper chain of Fig.~\ref{fig:501}, at mode $\Mone$ we always have $\omega_{n}(t) \leq \sP_{n,11}$, and at mode $\Mtwo$ it is $\omega_{n}(t) \leq \pi_{n,\sON}$. A packet is served if and only if $\Mone$ is visited in the upper chain of Fig.~\ref{fig:501}. 

To upper bound throughput, we compare Fig.~\ref{fig:501} to the mode transition diagrams in Fig.~\ref{fig:401} that corresponds to a fictitious model for channel $n$. This fictitious channel has constant information state $\omega_{n}(t) = \sP_{n,11}$ whenever it is in mode $\mathsf{M1}$, and $\omega_{n}(t) = \pi_{n,\sON}$ whenever it is in $\mathsf{M2}$. In other words, when the fictitious channel $n$ is in mode $\mathsf{M1}$ (or $\mathsf{M2}$), it sets its current information state to be the best state possible when the corresponding real channel $n$ is in the same mode. It follows that, when both the real and the fictitious channel $n$ are served, the probabilities of transitions $\Mone \to \Mone$ and $\Mtwo \to \Mone$ in the upper chain of Fig.~\ref{fig:401} are greater than or equal to those in Fig.~\ref{fig:501}, respectively. In other words, the upper chain of Fig.~\ref{fig:401} is \emph{more likely} to go to mode $\mathsf{M1}$ and serve packets than that of Fig.~\ref{fig:501}. Therefore, intuitively, if we serve both the real and the fictitious channel $n$ in the same infinite sequence of time slots, the fictitious channel $n$ will yield higher throughput for all $n$. This observation is made precise by the next lemma.

\begin{lem} \label{lem:701}
Consider  two discrete-time Markov chains $\{X(t)\}$ and $\{Y(t)\}$ both with state space $\{0, 1\}$. Suppose $\{X(t)\}$ is stationary and ergodic with transition probability matrix
\[
\bm{\sP} = \begin{bmatrix}
\sP_{00} & \sP_{01} \\ \sP_{10} & \sP_{11}
\end{bmatrix},
\]
and $\{Y(t)\}$ is non-stationary with
\[
\bm{\sQ}(t) = \begin{bmatrix}
\sQ_{00}(t) & \sQ_{01}(t) \\ \sQ_{10}(t) & \sQ_{11}(t)
\end{bmatrix}.
\]
Assume $\sP_{01} \geq \sQ_{01}(t)$ and $\sP_{11} \geq \sQ_{11}(t)$ for all $t$.  In $\{X(t)\}$, let $\pi_{X}(1)$ denote the stationary probability of state $1$; $\pi_{X}(1) = \sP_{01}/(\sP_{01}+\sP_{10})$. In $\{Y(t)\}$, define
\[
\pi_{Y}(1) \triangleq \limsup_{T\to\infty} \frac{1}{T} \sum_{t=0}^{T-1} Y(t)
\]
as the limiting fraction of time $\{Y(t)\}$ stays at state $1$.
Then we have $\pi_{X}(1) \geq \pi_{Y}(1)$.
\end{lem}

\begin{IEEEproof}[Proof of Lemma~\ref{lem:701}]
Given in Appendix~\ref{pf:107}.
\end{IEEEproof}

We note that executing a scheduling policy in the network is to generate a sequence of channel selection decisions. By Lemma~\ref{lem:701}, if we apply the same sequence of channel selection decisions of some scheduling policy to the set of fictitious channels,  we will get higher throughput on every channel. A direct consequence of this is that the maximum sum throughput over the fictitious channels is greater than or equal to that over the real channels.

\begin{lem} \label{lem:401}
The maximum sum throughput over the set of fictitious channels is no more than
\[
\max_{n\in\{1, 2, \ldots, N\}} \{c_{n,\infty}\}, \quad c_{n,\infty} \triangleq \frac{\sP_{n, 01}}{x_{n} \sP_{n, 10}+ \sP_{n, 01}}.
\]
\end{lem}
\begin{IEEEproof}[Proof of Lemma~\ref{lem:401}]
We note that finding the maximum sum throughput over fictitious channels in Fig.~\ref{fig:401} is equivalent to solving a multi-armed bandit problem~\cite{Git89book} with each channel acting as an arm (see Fig.~\ref{fig:401} and note that a  channel can change mode only when it is served), and one unit of reward is earned if a packet is delivered (recall that a packet is served if and only if mode $\Mone$ is visited in the upper chain of Fig.~\ref{fig:401}). The optimal solution to the multi-armed bandit system is to always play the arm (channel) with the largest average reward (throughput). The average reward  over channel $n$ is equal to the stationary probability of mode $\Mone$ in the upper chain of Fig.~\ref{fig:401}, which is
\[
\frac{\pi_{n, \sON}}{\sP_{n, 10} + \pi_{n, \sON}} = \frac{\sP_{n, 01}}{x_{n} \sP_{n, 10} + \sP_{n, 01}}.
\]
This finishes the proof.
\end{IEEEproof}

Together with the fact that throughput over any real channel $n$ cannot exceed its stationary $\sON$ probability $\pi_{n,\sON}$, we have constructed an outer bound on the network capacity region $\Lambda$ (the proof follows the above discussions and thus is omitted).

\begin{thm} (Generalized Outer Capacity Bound): \label{thm:401}
Any supportable throughput vector $\blambda = (\lambda_{1}, \lambda_{2}, \ldots, \lambda_{N})$ necessarily satisfies
\[
\begin{split}
\lambda_{n} &\leq \pi_{n, \sON}, \quad \text{for all $n\in\{1, 2, \ldots, N\}$,} \\
\sum_{n=1}^{N} \lambda_{n} &\leq \max_{n\in\{1, 2, \ldots, N\}} \left\{ c_{n,\infty} \right\} \\
&=\max_{n\in\{1, 2, \ldots, N\}} \left\{ \frac{\sP_{n,01}}{x_{n}\sP_{n,10}+\sP_{n,01}}\right\}.
\end{split}
\]
These $(N+1)$ hyperplanes create an outer capacity bound $\Lambda_{\text{out}}$ on $\Lambda$.
\end{thm}

\begin{cor}[Outer Capacity Bound for Symmetric Channels] \label{cor:j101}
In symmetric channels with $\bm{\sP}_{n} = \bm{\sP}$, $c_{n,\infty} = c_{\infty}$, and $\pi_{n,\sON} = \pi_{\sON}$ for all $n$, we have
\begin{equation}
\Lambda_{\text{out}} =
\left\{
\blambda \geq \bm{0} \mid
\sum_{n=1}^{N} \lambda_{n} \leq c_{\infty},\,\lambda_{n} \leq \pi_{\sON}\text{ for $1\leq n\leq N$}
\right\}, \label{eq:129}
\end{equation}
where $\geq$ is taken entrywise.
\end{cor}

We note that Lemma~\ref{lem:108} in Section~\ref{sec:pointproximity} directly follows Corollary~\ref{cor:j101}.


\subsection{A Two-User Example on Symmetric Channels} \label{sec:twousereg}
\begin{figure}[htbp]
\centering
\includegraphics[width=3in]{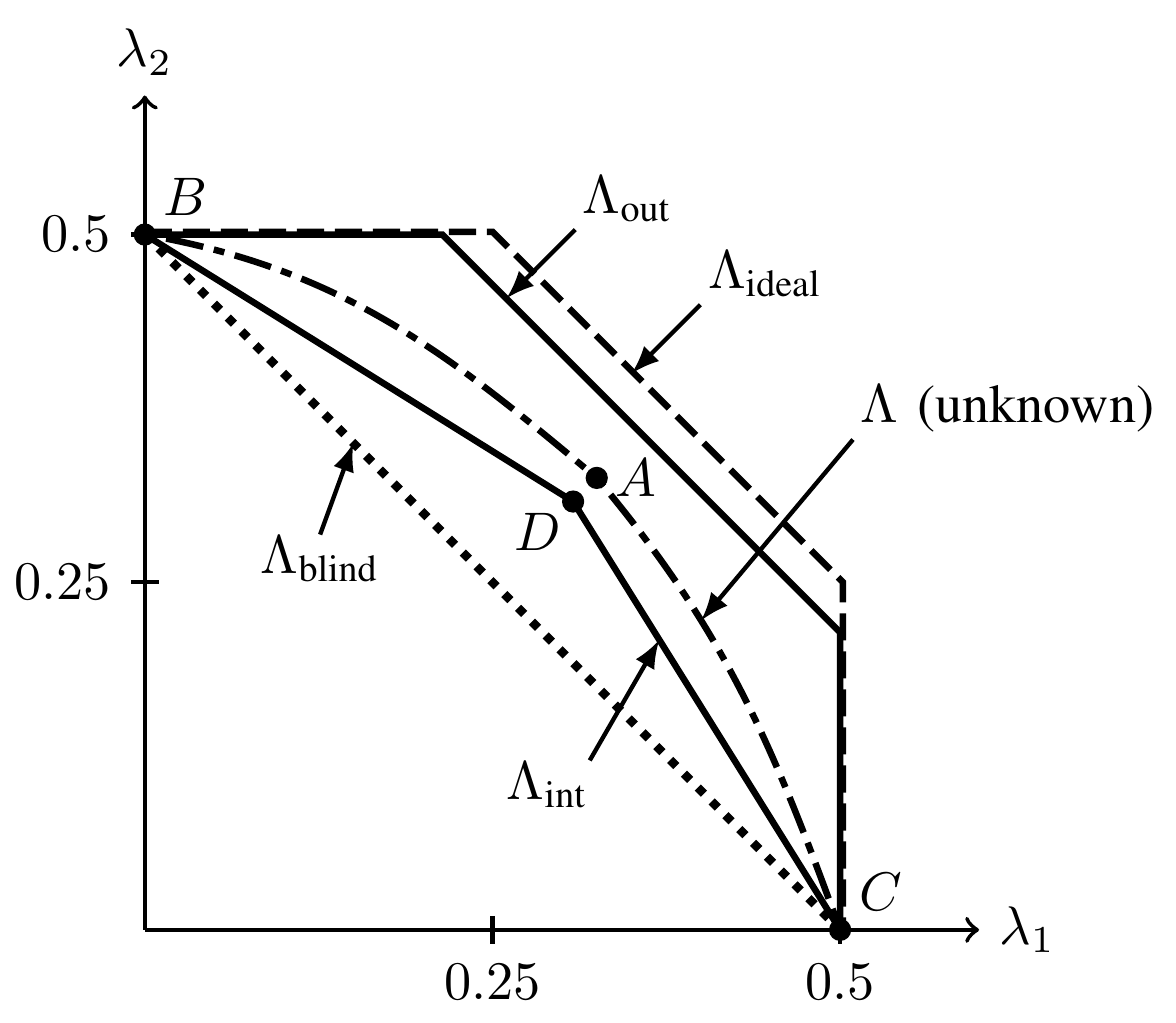}
\caption{Comparison of rate regions under different assumptions.}
\label{fig:101}
\end{figure}

Here we consider a two-user example on symmetric channels. For simplicity we will drop the subscript $n$ in notations. From Corollary~\ref{cor:j101}, we have the outer bound
\[
\Lambda_{\text{out}} = \Set{
\cvec{\lambda_{1} \\ \lambda_{2}} |
	\begin{gathered}[c]
	 0\leq \lambda_{n}\leq \sP_{01}/x,\text{ for $1\leq n\leq 2$}, \\
	 \lambda_{1} + \lambda_{2} \leq \sP_{01} / (x\sP_{10}+\sP_{01}), \\
	 x = \sP_{01} + \sP_{10}
	 \end{gathered}
}.
\]
For the inner bound $\Lambda_{\text{int}}$, we note that policy $\mathsf{RandRR}$ can execute three round robin policies $\mathsf{RR}(\bphi)$ for $\bphi \in \Phi = \{(1,1), (0,1), (1,0)\}$. From Corollary~\ref{cor:202}, we have
\[
\Lambda_{\text{int}} = \Set{
\cvec{\lambda_{1} \\ \lambda_{2}} |
	\begin{gathered}[c]
		0\leq \lambda_{n}\leq \mu_{n}, \text{ for $1\leq n\leq 2$,} \\
		\cvec{\mu_{1}\\ \mu_{2}}\in\conv{
			\left\{
			\cvec{c_{2}/2 \\ c_{2}/2},
			\cvec{c_{1} \\ 0},
			\cvec{0 \\ c_{1}}
			\right\}
		}
	\end{gathered}
}.
\]
Under the special case $\sP_{01} = \sP_{10} = 0.2$, the two bounds $\lambda_{\text{int}}$ and $\Lambda_{\text{out}}$ are shown in Fig.~\ref{fig:101}.

In Fig.~\ref{fig:101}, we also compare $\Lambda_{\text{int}}$ and $\Lambda_{\text{out}}$ with other rate regions. Set $\Lambda_{\text{ideal}}$ is the ideal capacity region when instantaneous channel states are known without causing any (timing) overhead~\cite{TaE93}. Next, it is shown in~\cite{ZKL08} that the maximum sum throughput in this network is achieved at point $A = (0.325, 0.325)$. The (unknown) network capacity region $\Lambda$ is bounded between $\Lambda_{\text{int}}$ and $\Lambda_{\text{out}}$, and has boundary points $B$, $A$, and $C$. Since the boundary of $\Lambda$ is a concave curve connecting $B$, $A$, and $C$, we envision that $\Lambda$ shall contain but be very close to $\Lambda_{\text{int}}$.

Finally, the rate region $\Lambda_{\text{blind}}$ is rendered by completely neglecting channel memory and treating the channels as i.i.d. over slots~\cite{LaN10}. We observe the throughput gain $\Lambda_{\text{int}} \setminus \Lambda_{\text{blind}}$, as much as $23\%$ in this example, is achieved by incorporating channel memory. In general, if channels are symmetric and treated as i.i.d. over slots, the maximum sum throughput in the network is $\pi_{\sON} = c_{1}$. Then the maximum throughput gain of $\mathsf{RandRR}$ using channel memory is $c_{N} - c_{1}$, which as $N\to\infty$ converges to
\[
c_{\infty} - c_{1} = \frac{\sP_{01}}{x \sP_{10} + \sP_{01}} - \frac{\sP_{01}}{\sP_{01}+\sP_{10}},
\]
which is controlled by the factor $x = \sP_{01}+\sP_{10}$.

\subsection{A Heuristically Tighter Inner Bound}
\begin{figure}[htbp]
\centering
\includegraphics[width=3in]{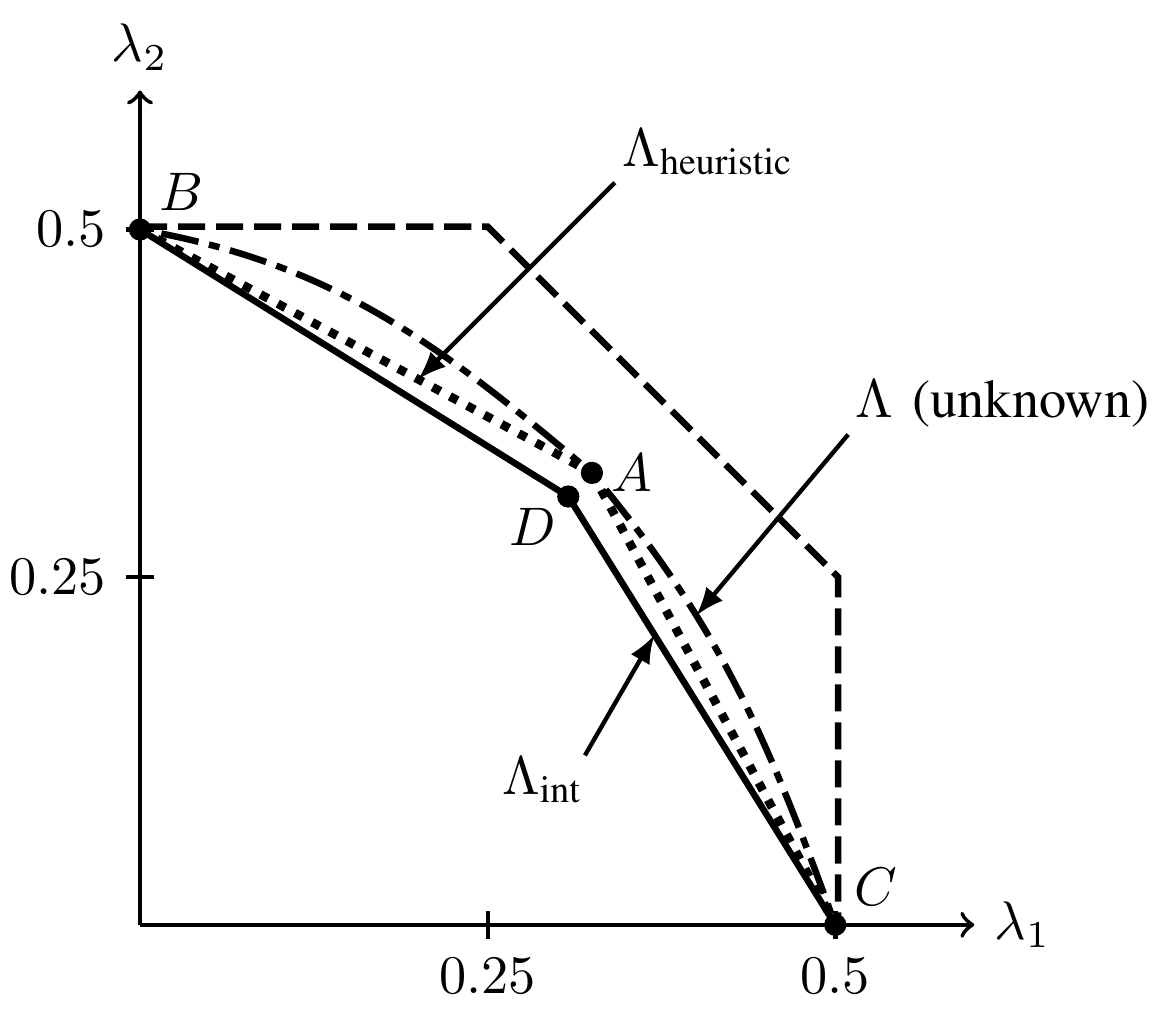}
\caption{Comparison of our inner bound $\Lambda_{\text{int}}$, the unknown network capacity region $\Lambda$, and a heuristically better inner bound $\Lambda_{\text{heuristic}}$.}
\label{fig:j101}
\end{figure}

It is shown in~\cite{ALJ09} that the following policy maximizes the sum throughput in a symmetric network:
\begin{quote}
\textit{Serve channels in a circular order, where on each channel keep transmitting data packets until a NACK is received}.
\end{quote}
In the above two-user example, this policy achieves throughput vector $A$ in Fig.~\ref{fig:j101}. 
If we replace our round robin policy $\mathsf{RR(\bphi)}$ by this one, heuristically we are able to construct a tighter inner capacity bound. For example,  we can support the tighter inner bound $\Lambda_{\text{heuristic}}$ in Fig.~\ref{fig:j101} by appropriate time sharing among the above policy that serves different subsets of channels. However, we note that this approach is difficult to analyze because the $\{L_{kn}\}$ process (see~\eqref{eq:j108}) forms a high-order Markov chain.  Yet, our inner bound $\Lambda_{\text{int}}$ provides a good throughput guarantee for this class of heuristic policies.

\section{Proximity of the Inner Bound to the True Capacity Region --- Symmetric Case} \label{sec:801}
Next we bound the closeness of the boundaries of  $\Lambda_{\text{int}}$ and $\Lambda$ in the case of symmetric channels. In Section~\ref{sec:pointproximity}, by choosing $M=N$, we have provided such analysis for the boundary point in the direction $(1, 1, \ldots, 1)$. Here we generalize to all boundary points. Define
\[
\mathcal{V} \triangleq \Set{ (v_{1}, v_{2}, \ldots, v_{N}) | 
	\begin{gathered}[c]
		v_{n}\geq 0 \text{ for $1\leq n\leq N$,} \\
		v_{n}>0 \text{ for at least one $n$}
	\end{gathered}
}
\]
as a set of directional vectors. For any  $\bv \in \mathcal{V}$, let $\blambda^{\text{int}} = (\lambda^{\text{int}}_{1}, \lambda^{\text{int}}_{2}, \ldots, \lambda^{\text{int}}_{N})$ and $\blambda^{\text{out}} = (\lambda^{\text{out}}_{1}, \lambda^{\text{out}}_{2}, \ldots, \lambda^{\text{out}}_{N})$ be the boundary point of $\Lambda_{\text{int}}$ and $\Lambda_{\text{out}}$ in the direction of $\bv$, respectively. It is useful to compute $\sum_{n=1}^{N} (\lambda_{n}^{\text{out}} - \lambda_{n}^{\text{int}})$, because it upper bounds the loss of the sum throughput of $\Lambda_{\text{int}}$ from $\Lambda$ in the direction of $\bv$.\footnote{Note that $\sum_{n=1}^{N} (\lambda_{n}^{\text{out}} - \lambda_{n}^{\text{int}})$ also bounds the closeness between $\Lambda_{\text{out}}$ and $\Lambda$.} We note that computing $\blambda^{\text{int}}$ in an arbitrary direction is difficult. Thus we will find an upper bound on  $\sum_{n=1}^{N} (\lambda_{n}^{\text{out}} - \lambda_{n}^{\text{int}})$.

\subsection{Preliminary}

To have more intuitions on $\Lambda_{\text{int}}$, we start with a toy example of  $N=3$ users. We are interested in the boundary point of $\Lambda_{\text{int}}$ in the direction of $\bv = (1,2,1)$. Consider two $\mathsf{RandRR}$-type policies $\psi_{1}$ and $\psi_{2}$ defined as follows. 
\begin{align*}
&\text{For $\psi_{1}$, choose } \begin{cases}
\bphi^{1} = (1, 0, 0) & \text{with prob. $1/4$} \\
\bphi^{2} = (0, 1, 0) & \text{with prob. $1/2$} \\
\bphi^{3} = (0, 0, 1) & \text{with prob. $1/4$}
\end{cases} \\
&\text{For $\psi_{2}$, choose } \begin{cases}
\bphi^{4} = (1, 1, 0) & \text{with prob. $1/2$} \\
\bphi^{5} = (0, 1, 1) & \text{with prob. $1/2$} \\
\end{cases}
\end{align*}
Both $\psi_{1}$ and $\psi_{2}$ support data rates in the direction of $(1,2,1)$. However, using the analysis of Lemma~\ref{lem:203} and Theorem~\ref{thm:205}, we know $\psi_{1}$ supports throughput vector
\[
\frac{1}{4} \cvec{c_{1} \\ 0 \\ 0} + \frac{1}{2} \cvec{0 \\ c_{1} \\ 0} + \frac{1}{4} \cvec{0 \\ 0 \\ c_{1}} = \frac{c_{1}}{4} \cvec{1 \\ 2  \\ 1},
\]
while $\psi_{2}$ supports
\[
\frac{1}{2} \cvec{c_{2}/2 \\ c_{2}/2 \\ 0} + \frac{1}{2} \cvec{0 \\ c_{2}/2 \\ c_{2}/2}
= \frac{c_{2}}{4} \cvec{1 \\ 2  \\ 1} \geq \frac{c_{1}}{4} \cvec{1 \\ 2  \\ 1},
\]
where $c_{1}$ and $c_{2}$ are defined in~\eqref{eq:j101}. We see that $\psi_{2}$ achieves data rates closer than $\psi_{1}$ does to the boundary of $\Lambda_{\text{int}}$. It is because every sub-policy of $\psi_{2}$, namely $\mathsf{RR}(\bphi^{4})$ and $\mathsf{RR}(\bphi^{5})$, supports sum throughput $c_{2}$ (by Lemma~\ref{lem:203}), where those of $\psi_{1}$ only support $c_{1}$. In other words, policy $\psi_{2}$ has better \emph{multiuser diversity gain} than $\psi_{1}$ does. This example suggests that we can find a good lower bound on $\blambda^{\text{int}}$ by exploring to what extent the multiuser diversity  can be exploited.  We start with the following definition.

\begin{defn} \label{defn:101}
For any $\bv\in\mathcal{V}$, we say $\bv$  is \emph{$d$-user diverse} if $\bv$ can be written as a \emph{positive} combination of vectors in  $\Phi_{d}$, where $\Phi_{d}$ denotes the set of $N$-dimensional binary vectors having $d$ entries be $1$. Define
\[
d(\bv) \triangleq \max_{1\leq d\leq N}\{ d \mid \text{$\bv$ is $d$-user diverse}\},
\]
and we shall say $\bv$ is \emph{maximally $d(\bv)$-user diverse}.
\end{defn}

The notion of $d(\bv)$ is well-defined because every $\bv$ must be $1$-user diverse.\footnote{The set $\Phi_{1} = \{\bm{e}_{1}, \bm{e}_{2}, \ldots, \bm{e}_{N}\}$ is the collection of unit coordinate vectors where $\bm{e}_{n}$ has its $n$th entry be $1$ and $0$ otherwise. Any vector $\bv\in\mathcal{V}$, $\bv = (v_{1}, v_{2}, \ldots, v_{N})$, can be written as $\bv = \sum_{v_{n} > 0} v_{n} \bm{e}_{n}$.}  Definition~\ref{defn:101} is the most useful to us through the next lemma.
\begin{lem} \label{lem:110}
The boundary point of $\Lambda_{\text{int}}$ in the direction of $\bv\in\mathcal{V}$ has sum throughput at least $c_{d(\bv)}$, where
\[
c_{d(\bv)} \triangleq \frac{
	\sP_{01}(1-(1-x)^{d(\bv)})
}{
	x\,\sP_{10} + \sP_{01}(1-(1-x)^{d(\bv)})
}, \quad x \triangleq \sP_{01}+\sP_{10}.
\]
\end{lem}
\begin{IEEEproof}[Proof of Lemma~\ref{lem:110}]
If direction $\bv$ can be written as a positive weighted sum of vectors in $\Phi_{d(\bv)}$, we can normalize the weights, and use  the new weights as probabilities to randomly mix $\mathsf{RR}(\bphi)$ policies for all $\bphi\in\Phi_{d(\bv)}$. This way we achieve sum throughput $c_{d(\bv)}$ in every transmission round, and overall the throughput vector will be in the direction of $\bv$. Therefore the result follows. For details,  see Appendix~\ref{pf:102}.
\end{IEEEproof}

Fig.~\ref{fig:j103} provides an example of Lemma~\ref{lem:110} in the two-user symmetric system in Section~\ref{sec:twousereg}.
\begin{figure}[htbp]
\centering
\includegraphics[width=2.5in]{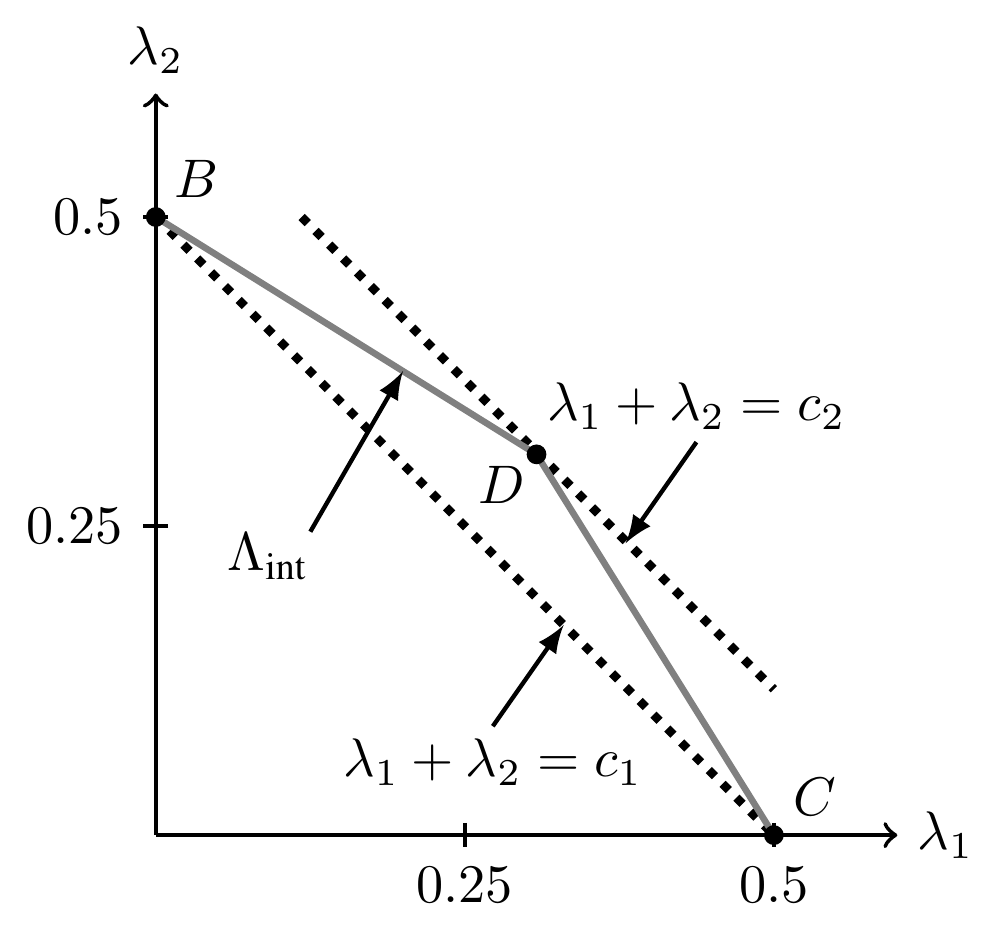}
\caption{An example for Lemma~\ref{lem:110} in the two-user symmetric network. Point $B$ and $C$ achieve  sum throughput $c_{1} = \pi_{\sON} = 0.5$, and the sum throughput at $D$ is $c_{2} \approx 0.615$. Any other boundary point of $\Lambda_{\text{int}}$ has sum throughput between $c_{1}$ and $c_{2}$.}
\label{fig:j103}
\end{figure}
We observe that direction $(1, 1)$, the one that passes point $D$ in Fig.~\ref{fig:j103}, is the only direction that is maximally $2$-user diverse. The sum throughput $c_{2}$ is achieved at $D$. For all the other directions, they are maximally $1$-user diverse and, from Fig.~\ref{fig:j103}, only sum throughput $c_{1}$ is guaranteed along those directions.  In general, geometrically we can show that a maximally $d$-user diverse vector, say $\bv_{d}$, forms a smaller angle with the all-$1$ vector $(1, 1, \ldots, 1)$ than a maximally $d'$-user diverse vector, say $\bv_{d'}$, does if $d' < d$. In other words, data rates along $\bv_{d}$ are \emph{more balanced} than those along $\bv_{d'}$. Lemma~\ref{lem:110} states that we guarantee to support higher sum throughput if the  user traffic is more balanced.

\subsection{Proximity Analysis}
We use the notion of $d(\bv)$ to upper bound $\sum_{n=1}^{N} (\lambda_{n}^{\text{out}} - \lambda_{n}^{\text{int}})$ in any direction $\bv\in\mathcal{V}$.  Let $\blambda^{\text{out}} = \theta \blambda^{\text{int}}$ (i.e., $\lambda_{n}^{\text{out}} = \theta \lambda_{n}^{\text{int}}$ for all $n$) for some $\theta \geq 1$. By~\eqref{eq:129}, the boundary of $\Lambda_{\text{out}}$ is characterized by the interaction of the $(N+1)$ hyperplanes $\sum_{n=1}^{N} \lambda_{n} = c_{\infty}$ and $\lambda_{n} = \pi_{\sON}$ for each $n\in\{1, 2, \ldots, N\}$. Specifically, in any given direction, if we consider the cross points on all the hyperplanes in that direction, the boundary point $\blambda^{\text{out}}$ is the one closest to the origin. We do not know which hyperplane $\blambda^{\text{out}}$ is on, and thus need to consider all $(N+1)$ cases. If $\blambda^{\text{out}}$ is on the plane $\sum_{n=1}^{N} \lambda_{n} = c_{\infty}$, i.e., $\sum_{n=1}^{N} \lambda_{n}^{\text{out}} = c_{\infty}$, we get
\[
\sum_{n=1}^{N} (\lambda_{n}^{\text{out}} - \lambda_{n}^{\text{int}}) \stackrel{(a)}{\leq} c_{\infty} - c_{d(\bv)} \stackrel{(b)}{\leq} c_{\infty}(1-x)^{d(\bv)},
\]
where (a) is by Lemma~\ref{lem:110} and (b) is by~\eqref{eq:222}. If $\blambda^{\text{out}}$ is on the plane $\lambda_{n} = \pi_{\sON}$ for some $n$, then $\theta = \pi_{\sON} / \lambda^{\text{int}}_{n}$. It follows
\[
\sum_{n=1}^{N} (\lambda_{n}^{\text{out}} - \lambda_{n}^{\text{int}}) = (\theta-1) \sum_{n=1}^{N} \lambda_{n}^{\text{int}} \leq \left(\frac{\pi_{\sON}}{\lambda^{\text{int}}_{n}}-1\right)c_{\infty}.
\]
The above discussions lead  to the next lemma.
\begin{lem}\label{lem:111}
The loss of the sum throughput of $\Lambda_{\text{int}}$ from $\Lambda$ in the direction of $\bv$ is upper bounded by
\begin{align}
& \min\left[ c_{\infty}(1-x)^{d(\bv)},\,\min_{1\leq n\leq N} \left\{ \left(\frac{\pi_{\sON}}{\lambda^{\text{int}}_{n}}-1\right)c_{\infty} \right\}\right] \notag \\
&\quad = c_{\infty} \min\left[  (1-x)^{d(\bv)},\,\frac{\pi_{\sON}}{\max_{1\leq n\leq N} \{\lambda^{\text{int}}_{n}\}} - 1 \right].  \label{eq:133}
\end{align}
\end{lem}

Lemma~\ref{lem:111} shows that, if data rates are more balanced, namely, have a larger $d(\bv)$, the sum throughput loss is dominated by the first term in the minimum of~\eqref{eq:133} and decreases to $0$ geometrically fast with $d(\bv)$. If data rates are biased toward a particular user, the second term in the minimum of~\eqref{eq:133} captures the throughput loss, which goes to $0$ as the rate of the favored user goes to the single-user capacity $\pi_{\sON}$.

\section{Throughput-Achieving Queue-dependent Round Robin Policy} \label{sec:501}
Let $a_{n}(t)$, for $1\leq n\leq N$, be the number of exogenous packet arrivals destined for user $n$ in slot $t$.  Suppose $a_{n}(t)$ are independent across users, i.i.d. over slots  with rate $\expect{a_{n}(t)} = \lambda_{n}$, and $a_{n}(t)$ is bounded with $0\leq a_{n}(t) \leq A_{\text{max}}$, where $A_{\text{max}}$ is a finite integer.  Let $U_{n}(t)$ be the backlog of user-$n$ packets queued at the base station at time $t$. Define $\bU(t) \triangleq (U_{1}(t), U_{2}(t), \ldots, U_{N}(t))$ and suppose $U_{n}(0) =0$ for all $n$. The queue process $\{U_{n}(t)\}$ evolves as 
\begin{equation} \label{eq:212}
U_{n}(t+1) = \max\left[U_{n}(t) - \mu_{n}(s_{n}(t), t), 0\right] + a_{n}(t),
\end{equation}
where $\mu_{n}(s_{n}(t), t)\in\{0, 1\}$ is the service rate allocated to user $n$ in slot $t$. We have $\mu_{n}(s_{n}(t), t) = 1$ if user $n$ is served and $s_{n}(t)=\sON$, and $0$ otherwise. In the rest of the paper we drop $s_{n}(t)$ in $\mu_{n}(s_{n}(t), t)$ and use $\mu_{n}(t)$ for notational simplicity. We say the network is (strongly) stable if
\[
\limsup_{t\to\infty} \frac{1}{t} \sum_{\tau=0}^{t-1} \sum_{n=1}^{N} \expect{U_{n}(\tau)} < \infty.
\]

Consider a rate vector $\blambda$ interior to the inner capacity region bound $\Lambda_{\text{int}}$ given in Theorem~\ref{thm:205}. Namely, there exists  an $\epsilon > 0$ and a probability distribution $\{\beta_{\bphi}\}_{\bphi\in\Phi}$ such that 
\begin{equation} \label{eq:209}
\lambda_{n} + \epsilon < \sum_{\bphi\in\Phi} \beta_{\bphi} \eta_{n}^{\bphi}, \quad \text{for all $1\leq n\leq N$},
\end{equation}
where $\eta_{n}^{\bphi}$ is defined in~\eqref{eq:509}. By Theorem~\ref{thm:205}, there exists a $\mathsf{RandRR}$ policy that yields service rates equal to the right-side of~\eqref{eq:209} and thus stabilizes the network with arrival rate vector $\blambda$~\cite[Lemma $3.6$]{GNT06}. The existence of this policy is useful and we shall denote it by $\mathsf{RandRR}^{*}$. Recall that on each new scheduling round, the policy $\mathsf{RandRR}^{*}$ randomly picks a binary vector $\bphi$ using probabilities $\alpha_{\bphi}$ (defined over all of the $(2^{N}-1)$ subsets of users). The $M(\bphi)$ active users in $\bphi$ are served for one round by the round robin policy $\mathsf{RR}(\bphi)$, serving the least recently used users first. However, solving for the probabilities needed to implement the $\mathsf{RandRR}^{*}$ policy that yields~\eqref{eq:209} is intractable when $N$ is large, because we need to find $(2^{N}-1)$ unknown probabilities $\{\alpha_{\bphi}\}_{\bphi\in\Phi}$, compute $\{\beta_{\bphi}\}_{\bphi\in\Phi}$ from~\eqref{eq:j508}, and make~\eqref{eq:209} hold. Instead of probabilistically finding the vector $\bphi$ for the current round of scheduling,  we use the following simple \emph{queue-dependent} policy.

\underline{\bf{Queue-dependent Round Robin Policy} ($\mathsf{QRR}$)}:
\begin{enumerate}
\item Start with $t=0$.
\item At time $t$, observe the current queue backlog vector $\bU(t)$ and find the binary vector $\bphi(t)\in\Phi$ defined as\footnote{The vector $\bphi(t)$ is a queue-dependent decision and thus we should write $\bphi(\bU(t), t)$ as a function of $\bU(t)$. For simplicity we use $\bphi(t)$ instead.}
\begin{equation} \label{eq:j504}
\bphi(t) \triangleq \arg\max_{\bphi\in\Phi} f(\bU(t), \mathsf{RR}(\bphi)),
\end{equation}
where
\begin{align*}
&f(\bU(t), \mathsf{RR}(\bphi))  \\
& \triangleq \sum_{n: \phi_{n}=1} \left[ U_{n}(t) \expect{L_{1n}^{\bphi}-1} - \expect{L_{1n}^{\bphi}} \sum_{n=1}^{N} U_{n}(t)\lambda_{n} \right] 
\end{align*}
and $\expect{L_{1n}^{\bphi}} = 1 + \sP_{n,01}^{(M(\bphi))} / \sP_{n,10}$ from~\eqref{eq:j105}. Ties are broken arbitrarily. \label{item:5}
\item Run $\mathsf{RR}(\bphi(t))$ for one round of transmission. We emphasize that active channels in $\bphi$ are served in the least-recently-used order. After the round ends, go to Step~\ref{item:5}.
\end{enumerate}

The $\mathsf{QRR}$ policy is a frame-based algorithm similar to $\mathsf{RandRR}$, except that at the beginning of every transmission round the policy selection is no longer random but based on a queue-dependent rule. We note that $\mathsf{QRR}$ is a polynomial time algorithm because we can compute $\bphi(t)$ in~\eqref{eq:j504} in polynomial time with the following divide and conquer approach:
\begin{enumerate}
\item Partition the set $\Phi$ into subsets $\{\Phi_{1}, \ldots, \Phi_{N}\}$, where $\Phi_{M}$, $M\in\{1, \ldots, N\}$, is the set  of $N$-dimensional binary vectors having exactly $M$ entries be $1$.
\item For each $M\in\{1, \ldots, N\}$, find the maximizer of $f(\bU(t), \mathsf{RR}(\bphi))$ among vectors in $\Phi_{M}$. For each $\bphi\in\Phi_{M}$, we have
\begin{align*}
&f(\bU(t), \mathsf{RR}(\bphi)) = \\
& \sum_{n: \phi_{n}=1} \left[ U_{n}(t) \frac{\sP_{n,01}^{(M)}}{\sP_{n,10}} - \left(1+  \frac{\sP_{n,01}^{(M)}}{\sP_{n,10}} \right) \sum_{n=1}^{N} U_{n}(t)\lambda_{n}  \right],
\end{align*}
and the maximizer of $f(\bU(t), \mathsf{RR}(\bphi))$ is to \emph{activate} the $M$ channels that yield the $M$ largest summands of the above equation.

\item Obtain $\bphi(t)$ by comparing the maximizers from the above step for different values of $M$.
\end{enumerate}
The detailed implementation is as follows.

\underline{\bf{Polynomial time implementation of Step~\ref{item:5} of $\mathsf{QRR}$}}:
\begin{enumerate}
\item For each fixed $M\in\{1, \ldots, N\}$, we do the following:

Compute
\begin{equation} \label{eq:j109}
U_{n}(t) \frac{\sP_{n,01}^{(M)}}{\sP_{n,10}} - \left(1+  \frac{\sP_{n,01}^{(M)}}{\sP_{n,10}} \right) \sum_{n=1}^{N} U_{n}(t)\lambda_{n} 
\end{equation}
for all $n\in\{1, \ldots, N\}$. Sort these $N$ numbers and define the binary vector $\bphi^{M} = (\phi_{1}^{M}, \ldots, \phi_{N}^{M})$ such that  $\phi_{n}^{M}=1$ if the value~\eqref{eq:j109} of channel $n$ is among the $M$ largest, otherwise $\phi_{n}^{M} = 0$. Ties are broken arbitrarily. Let $\hat{f}(\bU(t), M)$ denote the sum of the $M$ largest values of~\eqref{eq:j109}.

\item Define $M(t) \triangleq \arg\max_{1\leq M\leq N} \hat{f}(\bU(t),M)$. Then we assign $\bphi(t) = \bphi^{M(t)}$.
\end{enumerate}

Using a novel variable-length frame-based Lyapunov analysis, we show in the next theorem that $\mathsf{QRR}$ stabilizes the network with any arrival rate vector $\blambda$ strictly within the inner capacity bound $\Lambda_{\text{int}}$.\footnote{In~\eqref{eq:j602} we show that as long as the queue backlog vector $\bU(t)$ is not identically zero the arrival rate vector $\blambda$ is interior to the inner capacity bound $\Lambda_{\text{int}}$, in Step~\ref{item:5} of the $\mathsf{QRR}$ policy we always have
$
\max_{\bphi\in\Phi} f(\bU(t), \mathsf{RR}(\bphi)) >0.
$
}
The idea  is that we compare $\mathsf{QRR}$ with the (unknown) policy $\mathsf{RandRR}^{*}$ that stabilizes $\blambda$. We show that, in every transmission round, $\mathsf{QRR}$ finds and executes a round robin policy $\mathsf{RR}(\bphi(t))$ that yields a larger negative drift on the queue backlogs than $\mathsf{RandRR}^{*}$ does in the current round. Therefore, $\mathsf{QRR}$ is stable.

\begin{thm} \label{thm:102}
For any data rate vector $\blambda$ interior to $\Lambda_{\text{int}}$, policy $\mathsf{QRR}$ strongly stabilizes the network.
\end{thm}

\begin{IEEEproof}[Proof of Theorem~\ref{thm:102}]
See Appendix~\ref{pf:106}.
\end{IEEEproof}

\section{Conclusion}
The network capacity of a wireless network is practically degraded by communication overhead. In this paper, we take a step forward by studying the fundamental achievable rate region when communication overhead is kept minimum, that is, when channel probing is not permitted. While solving the original problem is difficult, we construct an inner and an outer bound on the network capacity region, with the aid of channel memory. When channels are symmetric and the network serves a large number of users, we show the inner and outer bound are progressively tight when the data rates of different users are more balanced. We also derive a simple queue-dependent frame-based policy, as a function of packet arrival rates and channel statistics, and show that this policy stabilizes the network for any data rates strictly within the inner capacity bound.

Transmitting data without channel probing is one of the many options for communication over a wireless network. Practically each option may have pros and cons on criteria like the achievable throughput, power efficiency, implementation complexity, etc. In the future it is important to explore how to combine all possible options to push the practically achievable network capacity to the limit.  It is part of our future work to generalize the methodology and framework developed in this paper to more general cases, such as when limited probing is allowed and/or other QoS metrics such as energy consumption are considered. It will also be interesting to see how this framework  can be applied to solve new problems in opportunistic spectrum access in cognitive radio networks, in opportunistic scheduling with delayed/uncertain channel state information, and in restless bandit problems.

\appendices

\section{} \label{pf:105}
\begin{IEEEproof}[Proof of Lemma~\ref{lem:201}]
Initially, by~\eqref{eq:202} we have $\omega_{n}(0) = \pi_{n,\sON} \geq \sP_{n,01}^{(M)}$ for all $n$. Suppose the base station switches to channel $n$ at time $t$, and the last use of channel $n$ ends at slot $(t-k)$ for some $k<t$. In slot $(t-k)$, there are two possible cases:
\begin{enumerate}
\item Channel $n$ turns $\sOFF$,  and as a result the information state on slot $t$ is $\omega_{n}(t) = \sP_{n,01}^{(k)}$. Due to round robin, the other $(M-1)$ channels must have been used for at least one slot before $t$ after slot $(t-k)$, and thus $k\ge M$. By~\eqref{eq:202} we have $\omega_{n}(t) = \sP_{n,01}^{(k)} \geq \sP_{n,01}^{(M)}$.
\item Channel $n$ is $\sON$ and transmits a dummy packet. Thus $\omega_{n}(t) = \sP_{n,11}^{(k)}$. By~\eqref{eq:202} we have $\omega_{n}(t) = \sP_{n,11}^{(k)} \geq \sP_{n,01}^{(M)}$.
\end{enumerate}
\end{IEEEproof}

\section{} \label{pf:103}
\begin{IEEEproof}[Proof of Lemma~\ref{lem:106}]
At the beginning of a new round, suppose round robin policy $\mathsf{RR}(\bphi)$ is selected. We index the $M(\bphi)$ active channels in $\bphi$ as $(n_{1}, n_{2}, \ldots, n_{M(\bphi)})$, which is in the decreasing order of the time duration between their last use and the beginning of the current round. In other words, the last use of $n_{k}$ is earlier than that of $n_{k'}$ only if $k < k'$. Fix an active channel $n_{k}$. Then it suffices to show that when this channel is served in the current round, the time duration back to the end of its last service is at least $(M(\bphi)-1)$ slots (that this channel has information state no worse than $\sP_{n_{k},01}^{(M(\bphi))}$ then follows the same arguments in the proof of Lemma~\ref{lem:201}).

We partition the active channels in $\bphi$ other than $n_{k}$ into two sets $\mathcal{A} = \{n_{1}, n_{2}, \ldots, n_{k-1}\}$ and $\mathcal{B} = \{n_{k+1}, n_{k+2}, \ldots, n_{M(\bphi)}\}$. Then the last use of every channel in $\mathcal{B}$ occurs after the last use of $n_{k}$, and so channel $n_{k}$ has been idled for at least $\abs{\mathcal{B}}$ slots at the start of the current round. However, the policy in this round will serve all channels in $\mathcal{A}$ before serving $n_{k}$, taking at least one slot per channel, and so we wait at least additional $\abs{\mathcal{A}}$ slots before serving channel $n_{k}$. The total time that this channel has been idled is thus at least $\abs{\mathcal{A}} + \abs{\mathcal{B}} = M(\bphi)-1$.
\end{IEEEproof}

\section{} \label{pf:104}

\begin{IEEEproof}[Proof of Theorem~\ref{thm:205}]
Let $Z(t)$ denote the number of times Step~\ref{item:4} of $\mathsf{RandRR}$ is executed in $[0, t)$, in which we suppose vector $\bphi$ is selected $Z_{\bphi}(t)$ times. Define $t_{i}$, where $i\in\Z^{+}$, as the $(i+1)$th time instant a new vector $\bphi$ is selected. Assume $t_{0} = 0$, and thus the first selection occurs at time $0$. It follows that $Z(t_{i}^{-}) = i$, $Z(t_{i}) = i+1$, and the $i$th round of packet transmissions ends at time $t_{i}^{-}$. 

Fix a vector $\bphi$.  Within the time periods in which policy $\mathsf{RR}(\bphi)$ is executed, denote by $L_{kn}^{\bphi}$ the duration of the $k$th time the base station stays with channel $n$.   Then the time average throughput that policy $\mathsf{RR}(\bphi)$ yields on its active channel $n$ over $[0, t_{i})$ is
\begin{equation} \label{eq:115}
\frac{
\sum_{k=1}^{Z_{\bphi}(t_{i})} \left(L_{kn}^{\bphi} - 1\right)
}{
\sum_{\bphi\in\Phi} \sum_{k=1}^{Z_{\bphi}(t_{i})} \sum_{n: \phi_{n} = 1} L_{kn}^{\bphi}
}.
\end{equation}
For simplicity, here we focus on discrete time instants $\{t_{i}\}$ large enough so that $Z_{\bphi}(t_{i})>0$ for all $\bphi\in\Phi$ (so that the sums in~\eqref{eq:115} make sense). The generalization to arbitrary time $t$ can be done by incorporating fractional transmission rounds, which are amortized over time. Next, rewrite~\eqref{eq:115} as
\begin{equation} \label{eq:117}
\frac{
\sum_{k=1}^{Z_{\bphi}(t_{i})} \sum_{n: \phi_{n} = 1} L_{kn}^{\bphi}
}{
\sum_{\bphi\in\Phi} \sum_{k=1}^{Z_{\bphi}(t_{i})} \sum_{n: \phi_{n} = 1} L_{kn}^{\bphi}
}
\underbrace{
\frac{
\sum_{k=1}^{Z_{\bphi}(t_{i})}  \left(L_{kn}^{\bphi} - 1\right)
}{
\sum_{k=1}^{Z_{\bphi}(t_{i})} \sum_{n: \phi_{n} = 1} L_{kn}^{\bphi}
}
}_{(\ast)}.
\end{equation}
As $t\to\infty$, the second term $(\ast)$ of~\eqref{eq:117} satisfies
\begin{align*}
(\ast)
&=
\frac{
 \frac{1}{Z_{\bphi}(t_{i})} \sum_{k=1}^{Z_{\bphi}(t_{i})}  \left(L_{kn}^{\bphi} - 1\right)
}{
\sum_{n: \phi_{n} = 1} \frac{1}{Z_{\bphi}(t_{i})} \sum_{k=1}^{Z_{\bphi}(t_{i})}  L_{kn}^{\bphi}
} \\
&\stackrel{(a)}{\to}
\frac{
	 \expect{L_{1n}^{\bphi} - 1}
}{
	\sum_{n: \phi_{n} = 1} \expect{L_{1n}^{\bphi}}
} \stackrel{(b)}{=} \eta_{n}^{\bphi},
\end{align*}
where (a) is by the Law of Large Numbers (we have shown in Corollary~\ref{cor:201} that $L_{kn}^{\bphi}$ are i.i.d. for different $k$) and (b) by~\eqref{eq:509}.

Denote the first term of~\eqref{eq:117} by $\beta_{\bphi}(t_{i})$, where we note that $\beta_{\bphi}(t_{i}) \in [0, 1]$ for all $\bphi\in\Phi$ and $\sum_{\bphi\in\Phi} \beta_{\bphi}(t_{i}) = 1$.  We can rewrite $\beta_{\bphi}(t_{i})$ as
\[
\beta_{\bphi}(t_{i}) =
\frac{
	\left[ \frac{Z_{\bphi}(t_{i})}{Z(t_{i})} \right]
	\sum_{n: \phi_{n} = 1}
		\left[
			\frac{1}{Z_{\bphi}(t_{i})} \sum_{k=1}^{Z_{\bphi}(t_{i})} L_{kn}^{\bphi}
		\right]
}{
	\sum_{\bphi\in\Phi}
	\left[ \frac{Z_{\bphi}(t_{i})}{Z(t_{i})} \right]
	\sum_{n: \phi_{n} = 1}
		\left[
			\frac{1}{Z_{\bphi}(t_{i})} \sum_{k=1}^{Z_{\bphi}(t_{i})} L_{kn}^{\bphi}
		\right]
}.
\]
As $t\to\infty$, we have
\begin{equation} \label{eq:206}
\beta_{\bphi} \triangleq \lim_{i\to\infty} \beta_{\bphi}(t_{i}) = \frac{
	\alpha_{\bphi} \sum_{n: \phi_{n}=1} \expect{L_{1n}^{\bphi}}
}{
	\sum_{\bphi\in\Phi} \alpha_{\bphi} \sum_{n: \phi_{n}=1} \expect{L_{1n}^{\bphi}}
},
\end{equation}
where by the Law of Large Numbers we have
\[
\frac{Z_{\bphi}(t_{i})}{Z(t_{i})} \to \alpha_{\bphi},\quad \frac{1}{Z_{\bphi}(t_{i})} \sum_{k=1}^{Z_{\bphi}(t_{i})} L_{kn}^{\bphi} \to  \expect{L_{1n}^{\bphi}}.
\]
From~\eqref{eq:115}\eqref{eq:117}\eqref{eq:206}, we have shown that the throughput contributed by policy $\mathsf{RR}(\bphi)$ on its active channel $n$ is $\beta_{\bphi} \eta_{n}^{\bphi}$. Consequently, $\mathsf{RandRR}$ parameterized by $\{\alpha_{\bphi}\}_{\bphi\in\Phi}$ supports any data rate vector $\blambda$ that is entrywise dominated by $\blambda \leq \sum_{\bphi\in\Phi} \beta_{\bphi} \bbeta^{\bphi}$,
where $\{\beta_{\bphi}\}_{\bphi\in\Phi}$ is defined in~\eqref{eq:206} and $\bbeta^{\bphi}$ in~\eqref{eq:509}.

The above analysis shows that every $\mathsf{RandRR}$ policy achieves a boundary point of $\Lambda_{\text{int}}$ defined in Theorem~\ref{thm:205}. Conversely, the next lemma, proved in Appendix~\ref{pf:101}, shows that every boundary point of $\Lambda_{\text{int}}$ is achievable by some $\mathsf{RandRR}$ policy, and the proof is complete.

\begin{lem} \label{lem:109}
For any probability distribution $\{\beta_{\bphi}\}_{\bphi\in\Phi}$, there exists  another probability distribution $\{\alpha_{\bphi}\}_{\bphi\in\Phi}$ that solves the linear system
\begin{equation} \label{eq:j508}
\beta_{\bphi} = \frac{
	\alpha_{\bphi} \sum_{n: \phi_{n}=1} \expect{L_{1n}^{\bphi}}
}{
	\sum_{\bphi\in\Phi} \alpha_{\bphi} \sum_{n: \phi_{n}=1} \expect{L_{1n}^{\bphi}}
},
\quad \text{for all $\bphi\in\Phi$.}
\end{equation}
\end{lem}
\end{IEEEproof}

\section{} \label{pf:101}
\begin{IEEEproof}[Proof of Lemma~\ref{lem:109}]
For any probability distribution $\{\beta_{\bphi}\}_{\bphi\in\Phi}$, we prove the lemma by inductively constructing the solution $\{\alpha_{\bphi}\}_{\bphi\in\Phi}$ to~\eqref{eq:j508}. The induction is on the cardinality of $\Phi$. Without loss of generality, we index elements in $\Phi$ by $\Phi = \{\bphi^{1}, \bphi^{2}, \ldots\}$, where $\bphi^{k} = (\phi_{1}^{k}, \ldots, \phi_{N}^{k})$. We define  $\chi_{k} \triangleq \sum_{n:\phi_{n}^{k}=1} \expect{L_{1n}^{\bphi^{k}}}$ and redefine $\beta_{\bphi^{k}} \triangleq \beta_{k}$ and $\alpha_{\bphi^{k}} \triangleq \alpha_{k}$. Then we can rewrite~\eqref{eq:j508} as
\begin{equation} \label{eq:207}
\beta_{k} = \frac{
	\alpha_{k} \chi_{k}
}{
	\sum_{1\leq k\leq \abs{\Phi}} \alpha_{k} \chi_{k},
}, 
\quad \text{for all $k\in\{1, 2, \ldots, \abs{\Phi}\}$.}
\end{equation}

We first note that $\Phi = \{\bphi^{1}\}$ is a degenerate case where $\beta_{1}$ and $\alpha_{1}$ must both be $1$. When $\Phi = \{\bphi^{1}, \bphi^{2}\}$, for any probability distribution $\{\beta_{1}, \beta_{2}\}$ with \emph{positive elements},\footnote{If one element of $\{\beta_{1}, \beta_{2}\}$ is zero, say $\beta_{2} = 0$, we can show necessarily $\alpha_{2}=0$ and it degenerates to the one-policy case $\Phi=\{\bphi^{1}\}$. Such degeneration happens in general cases. Thus in the rest of the proof we will only consider probability distributions that only have positive elements.} it is easy to show
\[
\alpha_{1} = \frac{
	\chi_{2} \beta_{1}
}{
	\chi_{1} \beta_{2} + \chi_{2} \beta_{1}
}, \quad \alpha_{2} = 1-\alpha_{1}.
\]
Let $\Phi = \{\bphi^{k}: 1\leq k\leq K\}$ for some $K\geq 2$. Assume that for any probability distribution $\{\beta_{k} >0: 1\leq k\leq K\}$ we can find $\{\alpha_{k}: 1\leq k\leq K\}$ that solves~\eqref{eq:207}. 

For the case $\Phi = \{\bphi^{k}: 1\leq k\leq K+1\}$ and any $\{\beta_{k}>0: 1\leq k\leq K+1\}$, we construct the solution $\{\alpha_{k}: 1\leq k\leq K+1\}$ to~\eqref{eq:206} as follows. Let $\{\gamma_{2}, \gamma_{3}, \ldots, \gamma_{K+1}\}$ be the solution to the linear system
\begin{equation} \label{eq:127}
\frac{
	\gamma_{k} \chi_{k}
}{
	\sum_{k=2}^{K+1} \gamma_{k} \chi_{k}
}
= \frac{
	\beta_{k}
}{
	\sum_{k=2}^{K+1} \beta_{k}
},
\quad 2\leq k\leq K+1.
\end{equation}
By the induction assumption, the set $\{\gamma_{2}, \gamma_{3}, \ldots, \gamma_{K+1}\}$ exists and satisfies $\gamma_{k} \in [0, 1]$ for $2\leq k\leq K+1$ and $\sum_{k=2}^{K+1} \gamma_{k} = 1$. Define
\begin{align}
\alpha_{1} &\triangleq
\frac{
	\beta_{1} \sum_{k=2}^{K+1} \gamma_{k} \chi_{k}
}{
	\chi_{1} (1-\beta_{1}) + \beta_{1} \sum_{k=2}^{K+1} \gamma_{k} \chi_{k}
} \label{eq:124} \\
\alpha_{k} &\triangleq (1-\alpha_{1}) \gamma_{k}, \quad 2\leq k\leq K+1. \label{eq:125}
\end{align}
It remains to show~\eqref{eq:124} and~\eqref{eq:125} are the desired solution.
It is easy to observe that $\alpha_{k}\in [0,1]$ for $1\leq k\leq K+1$, and
\[
\sum_{k=1}^{K+1} \alpha_{k} = \alpha_{1} + (1-\alpha_{1}) \sum_{k=2}^{K+1} \gamma_{k} = \alpha_{1} + (1-\alpha_{1}) = 1.
\]
By rearranging terms in~\eqref{eq:124} and using~\eqref{eq:125}, we have
\begin{equation} \label{eq:126}
\beta_{1} = \frac{
	\alpha_{1} \chi_{1}
}{
	\alpha_{1} \chi_{1} + \sum_{k=2}^{K+1} (1-\alpha_{1}) \gamma_{k} \chi_{k}
}
=
\frac{
	\alpha_{1} \chi_{1}
}{
	\sum_{k=1}^{K+1} \alpha_{k} \chi_{k}
}.
\end{equation}
For $2\leq k\leq K+1$,
\begin{align*}
\frac{
	\alpha_{k} \chi_{k}
}{
	\sum_{k=1}^{K+1} \alpha_{k} \chi_{k}
}
&=
\left[
\frac{
	\alpha_{k} \chi_{k}
}{
	\sum_{k=2}^{K+1} \alpha_{k} \chi_{k}
} \right] \left[
\frac{
	\sum_{k=2}^{K+1} \alpha_{k} \chi_{k}
}{
	\sum_{k=1}^{K+1} \alpha_{k} \chi_{k}
}
\right] \\
&\stackrel{(a)}{=} \!
\left[
\frac{
	(1-\alpha_{1})\gamma_{k} \chi_{k}
}{
	\sum_{k=2}^{K+1} (1-\alpha_{1})\gamma_{k} \chi_{k}
}
\right]\left[
1-\frac{
	\alpha_{1} \chi_{1}
}{
	\sum_{k=1}^{K+1} \alpha_{k} \chi_{k}
}
\right] \\
&\stackrel{(b)}{=}
\left[
\frac{
	\gamma_{k} \chi_{k}
}{
	\sum_{k=2}^{K+1} \gamma_{k} \chi_{k}
} \right]
(1-\beta_{1}) \\
&\stackrel{(c)}{=} \left(\frac{\beta_{k}}{\sum_{k=2}^{K+1} \beta_{k}} \right)  (1-\beta_{1}) \stackrel{(d)}{=} \beta_{k},
\end{align*}
where~(a) is by plugging in~\eqref{eq:125}, (b) uses~\eqref{eq:126},~(c) uses~\eqref{eq:127}, and~(d) is by $\sum_{k=1}^{K+1} \beta_{k} = 1$. The proof is complete.
\end{IEEEproof}

\section{} \label{pf:107}
\begin{IEEEproof}[Proof of Lemma~\ref{lem:701}]
Let $\cN_{1}(T) \subseteq \{0, 1, \ldots, T-1\}$ be the subset of time instants in which $Y(t) = 1$. Note that
$
\sum_{t=0}^{T-1} Y(t) = \abs{\cN_{1}(T)}.
$
For each $t\in\cN_{1}(T)$,  let $\1{1\to 0}(t)$ be an indicator function which is $1$ if $Y(t)$ transits from $1$ to $0$ at time $t$, and $0$ otherwise. We define $\cN_{0}(T)$ and $\1{0\to 1}(t)$ similarly.

In $\{0, 1, \ldots, T-1\}$, since  state transitions of $\{Y(t)\}$ from $1$ to $0$ and from $0$ to $1$ differ by at most $1$, we have
\begin{equation} \label{eq:702}
\abs{
\sum_{t\in\cN_{1}(T)} \1{1\to 0}(t) - \sum_{t\in\cN_{0}(T)} \1{0\to 1}(t)
} \leq 1,
\end{equation}
which is true for all $T$. Dividing~\eqref{eq:702} by $T$, we get
\begin{equation} \label{eq:703}
\abs{
\frac{1}{T}\sum_{t\in\cN_{1}(T)} \1{1\to 0}(t) - \frac{1}{T}\sum_{t\in\cN_{0}(T)} \1{0\to 1}(t)
} \leq \frac{1}{T}.
\end{equation}
Consider the subsequence $\{T_{k}\}$ such that
\begin{equation} \label{eq:704}
\lim_{k\to\infty} \frac{1}{T_{k}} \sum_{t=0}^{T_{k}-1} Y(t) = \pi_{Y}(1) = \lim_{k\to\infty} \frac{\abs{\cN_{1}(T_{k})}}{T_{k}}.
\end{equation}
Note that $\{T_{k}\}$ exists because $(1/T) \sum_{t=0}^{T-1} Y(t)$ is a bounded sequence indexed by integers $T$. Moreover, there exists a subsequence $\{T_{n}\}$ of $\{T_{k}\}$ so that each of the two averages in~\eqref{eq:703} has a limit point with respect to $\{T_{n}\}$, because they are bounded sequences, too. In the rest of the proof we will work on $\{T_{n}\}$, but we drop subscript $n$ for notational simplicity. Passing $T\to\infty$, we get from~\eqref{eq:703} that
\begin{equation} \label{eq:705}
\begin{split}
&
\underbrace{
\left(
\lim_{T\to\infty} \frac{\abs{\cN_{1}(T)}}{T} \right) }_{\stackrel{(a)}{=} \pi_{Y}(1)}
\underbrace{
\left(
\lim_{T\to\infty} \frac{1}{\abs{\cN_{1}(T)}} \sum_{t\in\cN_{1}(T)} \1{1\to 0}(t) \right)}_{\triangleq \beta}
\\
&
= 
\underbrace{
\left(
\lim_{T\to\infty} \frac{\abs{\cN_{0}(T)}}{T} \right)}_{\stackrel{(b)}{=} 1- \pi_{Y}(1)}
\underbrace{
\left(
\lim_{T\to\infty} \frac{1}{\abs{\cN_{0}(T)}} \sum_{t\in\cN_{0}(T)} \1{0\to 1}(t) \right)}_{\triangleq \gamma},
\end{split}
\end{equation}
where (a) is by~\eqref{eq:704} and (b) is by $\abs{\cN_{1}(T)} + \abs{\cN_{0}(T)} = T$. From~\eqref{eq:705} we get
\[
\pi_{Y}(1) = \frac{\gamma}{\beta + \gamma}.
\]
The next lemma, proved in Appendix~\ref{pf:108}, helps to show $\gamma \leq \sP_{01}$.

\begin{lem}[Stochastic coupling of random binary sequences] \label{lem:j101}
Let $\{I_{n}\}_{n=1}^{\infty}$ be an infinite sequence of binary random variables. Suppose for all $n\in\{1, 2, \ldots\}$ we have
\begin{equation} \label{eq:j110}
\prob{I_{n} = 1 \mid I_{1} = i_{1}, \ldots, I_{n-1} = i_{n-1}} \leq \sP_{01}
\end{equation}
for all possible values of $i_{1}, \ldots, i_{n-1}$. Then we can construct a new sequence $\{\hI_{n}\}_{n=1}^{\infty}$ of binary random variables that are i.i.d. with $\prob{\hI_{n} = 1} = \sP_{01}$ for all $n$ and satisfy $\hI_{n} \geq I_{n}$ for all $n$. Consequently, we have
\[
\limsup_{N\to\infty} \frac{1}{N} \sum_{n=1}^{N} I_{n} \leq \limsup_{N\to\infty} \frac{1}{N} \sum_{n=1}^{N} \hI_{n} = \sP_{01}.
\]
\end{lem}

To use Lemma~\ref{lem:j101} to prove $\gamma \leq \sP_{01}$, let $t_{n}$ denote the $n$th time $Y(t)=0$ and let $I_{n} = 1_{[0\to 1]}(t_{n})$. For simplicity assume $\{t_{n}\}$ is an infinite sequence so that state $0$ is visited infinitely often in $\{Y(t)\}$.  By the assumption that $\sQ_{01}(t) \leq \sP_{01}$ for all $t$, we know \eqref{eq:j110} holds. Therefore by Lemma~\ref{lem:j101} we have
\[
\gamma \leq \limsup_{N\to\infty} \frac{1}{N} \sum_{n=1}^{N} 1_{[0\to 1]}(t_{n}) \leq \sP_{01}.
\]

Similarly as Lemma~\ref{lem:j101}, we can show $\beta \geq \sP_{10}$ by stochastic coupling. Therefore
\[
\pi_{Y}(1) = \frac{\gamma}{\beta + \gamma} \leq \frac{\gamma}{\sP_{10} + \gamma} \leq \frac{\sP_{01}}{\sP_{01}+\sP_{10}} = \pi_{X}(1).
\]
\end{IEEEproof}

\section{} \label{pf:108}
\begin{IEEEproof}[Proof of Lemma~\ref{lem:j101}]
For simplicity, we assume 
\[
\prob{I_{n} = 0 \mid I_{1} = i_{1}, \ldots, I_{n-1} = i_{n-1}} > 0
\]
for all $n$ and all possible values of $i_{1}, \ldots, i_{n-1}$. For each $n\in\{1, 2, \ldots\}$, define $\hI_{n}$ as follows: If $I_{n} = 1$, define $\hI_{n} = 1$. If $I_{n} = 0$, observe the \emph{history} $I_{1}^{n-1} \triangleq (I_{1}, \ldots, I_{n-1})$ and independently choose $\hI_{n}$ as follows:
\begin{equation} \label{eq:j111}
\hI_{n} = \begin{cases}
			1 & \text{with prob. } \frac{\sP_{01} - \prob{I_{n}=1 \mid I_{1}^{n-1}}}{\prob{I_{n}=0 \mid I_{1}^{n-1}}} \\
			0 & \text{with prob. } 1 - \frac{\sP_{01} - \prob{I_{n}=1 \mid I_{1}^{n-1}}}{\prob{I_{n}=0 \mid I_{1}^{n-1}}}.
			\end{cases}
\end{equation}
The probabilities in~\eqref{eq:j111} are well-defined because $\sP_{01} \geq \prob{I_{n}=1 \mid I_{1}^{n-1}}$ by~\eqref{eq:j110}, and
\[
\sP_{01} \leq 1 = \prob{I_{n}=1 \mid I_{1}^{n-1}} +\prob{I_{n}=0 \mid I_{1}^{n-1}}
\]
and therefore 
\[
\sP_{01} - \prob{I_{n}=1 \mid I_{1}^{n-1}} \leq \prob{I_{n}=0 \mid I_{1}^{n-1}}.
\]

With the above definition of $\hI_{n}$, we have $\hI_{n} = 1$ whenever $I_{n} = 1$. Therefore $\hI_{n} \geq I_{n}$ for all $n$. Further, for any $n$ and any binary vector $i_{1}^{n-1}\triangleq (i_{1}, \ldots, i_{n-1})$, we have
\begin{equation} \label{eq:j112}
\begin{split}
& \prob{\hI_{n} = 1 \mid I_{1}^{n-1} = i_{1}^{n-1}} \\
& = \prob{I_{n} = 1 \mid I_{1}^{n-1} = i_{1}^{n-1}} + \prob{I_{n} = 0 \mid I_{1}^{n-1} = i_{1}^{n-1}}  \\
& \qquad  \times \frac{\sP_{01} - \prob{I_{n}=1 \mid I_{1}^{n-1} = i_{1}^{n-1}}}{\prob{I_{n}=0 \mid I_{1}^{n-1} = i_{1}^{n-1}}} = \sP_{01}.
\end{split}
\end{equation}
Therefore, for all $n$ we have
\begin{align*}
&\prob{\hI_{n} {}= 1} \\
& = \sum_{i_{1}^{n-1}} \prob{\hI_{n} = 1 \mid I_{1}^{n-1} = i_{1}^{n-1}}  \prob{I_{1}^{n-1} = i_{1}^{n-1}} = \sP_{01},
\end{align*}
and thus the $\hI_{n}$ variables are identically distributed. It remains to prove that they are independent.

Suppose components in $\hI_{1}^{n} \triangleq (\hI_{1}, \ldots, \hI_{n})$ are independent. We prove that components in $\hI_{1}^{n+1} = (\hI_{1}, \ldots, \hI_{n+1})$ are also independent.  For any binary vector $\hi_{1}^{n+1} \triangleq (\hi_{1}, \ldots, \hi_{n+1})$, since
\begin{align*}
& \prob{\hI_{1}^{n+1} = \hi_{1}^{n+1}} \\
& = \prob{\hI_{n+1} = \hi_{n+1} \mid \hI_{1}^{n} = \hi_{1}^{n}} \prob{\hI_{1}^{n} = \hi_{1}^{n}} \\
& = \prob{\hI_{n+1} = \hi_{n+1} \mid \hI_{1}^{n} = \hi_{1}^{n}} \prod_{k=1}^{n} \prob{\hI_{k} = \hi_{k}},
\end{align*}
it suffices to show
\[
\prob{\hI_{n+1} = 1 \mid \hI_{1}^{n} = \hi_{1}^{n}} = \prob{\hI_{n+1} = 1} = \sP_{01}.
\]
Indeed,
\begin{align*}
& \prob{\hI_{n+1} = 1 \mid \hI_{1}^{n} = \hi_{1}^{n}} \\
& = \sum_{i_{1}^{n}} \prob{\hI_{n+1} = 1 \mid I_{1}^{n} = i_{1}^{n}, \hI_{1}^{n} = \hi_{1}^{n}} \\
& \qquad \qquad \times \prob{I_{1}^{n} = i_{1}^{n} \mid \hI_{1}^{n} = \hi_{1}^{n}} \\
& = \sum_{i_{1}^{n}} \prob{\hI_{n+1} = 1 \mid I_{1}^{n} = i_{1}^{n}} \prob{I_{1}^{n} = i_{1}^{n} \mid \hI_{1}^{n} = \hi_{1}^{n}} \\
&\stackrel{(a)}{=} \sum_{i_{1}^{n}} \sP_{01} \prob{I_{1}^{n} = i_{1}^{n} \mid \hI_{1}^{n} = \hi_{1}^{n}} = \sP_{01},
\end{align*}
where (a) is by~\eqref{eq:j112}, and the proof is complete.
\end{IEEEproof}

\section{} \label{pf:102}
\begin{IEEEproof}[Proof of Lemma~\ref{lem:110}]
By definition of $d(\bv)$, there exists a nonempty subset $A \subseteq \Phi_{d(\bv)}$, and  for every $\bphi\in A$ a positive real number $\hat{\beta}_{\bphi}>0$,  such that
$
\bv = \sum_{\bphi\in A} \hat{\beta}_{\bphi} \bphi.
$
For each $\bphi\in A$, we have $M(\bphi) = d(\bv)$ and thus $c_{M(\bphi)} = c_{d(\bv)}$. Define
\[
\beta_{\bphi} \triangleq \frac{ \hat{\beta}_{\bphi} }{ \sum_{\bphi\in  A} \hat{\beta}_{\bphi} }
\]
for each $\bphi\in A$ and $\{\beta_{\bphi}\}_{\bphi\in A}$ is a probability distribution. Consider a $\mathsf{RandRR}$ policy that in every round selects $\bphi\in A$ with probability $\beta_{\bphi}$. By Lemma~\ref{lem:203}, this $\mathsf{RandRR}$ policy achieves throughput vector $\blambda = (\lambda_{1}, \ldots, \lambda_{N})$ that satisfies
\begin{align*}
\blambda = \sum_{\bphi\in A} \beta_{\bphi} \frac{c_{M(\bphi)}}{M(\bphi)} \bphi 
&= \frac{c_{d(\bv)}}{d(\bv)} \sum_{\bphi\in A} \frac{\hat{\beta}_{\bphi}}{ \sum_{\bphi\in A}\hat{\beta}_{\bphi}} \bphi  \\
&= \frac{c_{d(\bv)}}{d(\bv) \sum_{\bphi\in A} \hat{\beta}_{\bphi}} \sum_{\bphi\in A} \hat{\beta}_{\bphi} \bphi \\
&= \left( \frac{c_{d(\bv)}}{d(\bv) \sum_{\bphi\in A} \hat{\beta}_{\bphi}}\right)\,\bv,
\end{align*}
which is in the direction of $\bv$. In addition, the sum throughput
\[
\sum_{n=1}^{N} \lambda_{n} =  \sum_{\bphi \in A} \beta_{\bphi}  \frac{c_{M(\bphi)}}{M(\bphi)} \left(\sum_{n=1}^{N} \phi_{n} \right) = \sum_{\bphi \in A} \beta_{\bphi} c_{M(\bphi)} = c_{d(\bv)}
\]
is achieved.
\end{IEEEproof}

\section{} \label{pf:106}
\begin{IEEEproof}[Proof of Theorem~\ref{thm:102}]
({\bf A Related $\mathsf{RandRR}$ Policy})
For each randomized round robin policy $\mathsf{RandRR}$,  it is useful to consider a \emph{renewal reward process} where renewal epochs are defined as time instants at which $\mathsf{RandRR}$ starts a new round of transmission.\footnote{We note that the renewal reward process is defined solely with respect to $\mathsf{RandRR}$, and is only used to facilitate our analysis. At these renewal epochs, the state of the network, including the current queue state $\bU(t)$, does \emph{not} necessarily renew itself.}  Let $T$ denote the renewal period. We say one unit of reward is earned by a user if $\mathsf{RandRR}$ serves a packet to that user. Let $R_{n}$ denote the sum reward earned by user $n$ in one renewal period $T$, representing the number of successful transmissions user $n$ receives in one round of scheduling. Conditioning on the round robin policy $\mathsf{RR}(\bphi)$ chosen by $\mathsf{RandRR}$ for the current round of transmission, we have from Corollary~\ref{cor:201}:
\begin{gather}
\expect{T} = \sum_{\bphi\in\Phi} \alpha_{\bphi} \expect{T \mid \mathsf{RR}(\bphi)} \label{eq:502} \\
\expect{T\mid \mathsf{RR}(\bphi)} = \sum_{n:\phi_{n}=1} \expect{L_{1n}^{\bphi}} \label{eq:501},
\end{gather}
and for all $n\in\{1, 2, \ldots, N\}$,
\begin{gather}
\expect{R_{n}} = \sum_{\bphi\in\Phi} \alpha_{\bphi} \expect{R_{n} \mid \mathsf{RR}(\bphi)} \label{eq:504} \\
\expect{R_{n}\mid \mathsf{RR}(\bphi)}
=\begin{cases} 
\expect{L_{1n}^{\bphi} - 1} & \text{if $\phi_{n}=1$} \\
0 & \text{if $\phi_{n}=0$.}
	\end{cases} \label{eq:503}
\end{gather}

Consider the round robin policy $\mathsf{RR}((1, 1, \ldots, 1))$ that serves all $N$ channels in one round. We define $T_{\text{max}}$ as its renewal period. From Corollary~\ref{cor:201}, we know $\expect{T_{\text{max}}}<\infty$ and $\expect{(T_{\text{max}})^{2}}<\infty$. Further, for any $\mathsf{RandRR}$,  including using a $\mathsf{RR}(\bphi)$ policy in every round as special cases, we can show that $T_{\text{max}}$ is stochastically larger than the renewal period $T$, and $(T_{\text{max}})^{2}$ is stochastically larger than $T^{2}$.  It follows that
\begin{equation} \label{eq:307}
\expect{T} \leq \expect{T_{\text{max}}},\ \expect{T^{2}} \leq \expect{(T_{\text{max}})^{2}}.
\end{equation}

We have denoted by $\mathsf{RandRR}^{*}$ (in the discussion after~\eqref{eq:209}) the randomized round robin policy  that achieves a service rate vector strictly larger than the target arrival rate vector $\blambda$ entrywise. Let $T^{*}$ denote the  renewal period of $\mathsf{RandRR}^{*}$, and $R_{n}^{*}$ the sum reward (the number of successful transmissions) received by user $n$ over the renewal period $T^{*}$. Then we have
\begin{align}
\frac{\expect{R_{n}^{*}}}{\expect{T^{*}}} 
&\stackrel{(a)}{=} \frac{
	\sum_{\bphi\in\Phi} \alpha_{\bphi}\,\expect{R_{n}^{*} \mid \mathsf{RR}(\bphi)}
}{
	\sum_{\bphi\in\Phi} \alpha_{\bphi}\,\expect{T^{*} \mid \mathsf{RR}(\bphi)}
} \notag \\
&\stackrel{(b)}{=} \sum_{\bphi\in\Phi} \left( \frac{
	\alpha_{\bphi}
}{
	\sum_{\bphi\in\Phi} \alpha_{\bphi}\,\expect{T^{*} \mid \mathsf{RR}(\bphi)}
}\right) \expect{R_{n}^{*} \mid \mathsf{RR}(\bphi)} \notag \\
&= \sum_{\bphi\in\Phi} 
\underbrace{
	\frac{
		\alpha_{\bphi}\,\expect{T^{*} \mid \mathsf{RR}(\bphi)}
	}{
		\sum_{\bphi\in\Phi}\alpha_{\bphi}\,\expect{T^{*} \mid \mathsf{RR}(\bphi)}
	}
}_{(c)=\beta_{n}}\,
\underbrace{
	\frac{
		\expect{R_{n}^{*} \mid \mathsf{RR}(\bphi)}
	}{
		\expect{T^{*} \mid \mathsf{RR}(\bphi)}
}}_{(d)=\eta_{n}^{\bphi}} \notag \\
&= \sum_{\bphi\in\Phi} \beta_{\bphi} \eta_{n}^{\bphi} \stackrel{(e)}{>} \lambda_{n} +\epsilon, \label{eq:211}
\end{align}
where (a) is by~\eqref{eq:502}\eqref{eq:504}, (b) is by rearranging terms, (c) is by plugging \eqref{eq:501} into~\eqref{eq:j508}, (d) is by plugging~\eqref{eq:501} and~\eqref{eq:503} into~\eqref{eq:509} in Section~\ref{sec:802}, and (e) is by~\eqref{eq:209}. From~\eqref{eq:211} we get
\begin{equation} \label{eq:j501}
\expect{R_{n}^{*}} > (\lambda_{n} +\epsilon) \expect{T^{*}}, \quad\text{for all $n\in\{1, \ldots, N\}$.}
\end{equation}


({\bf Lyapunov Drift}) From~\eqref{eq:212},  in a frame of size $T$ (which is possibly random), we can show that for all $n$
\begin{equation} \label{eq:213}
U_{n}(t+T) \leq \max\left[ U_{n}(t) - \sum_{\tau=0}^{T-1} \mu_{n}(t+\tau), 0 \right] + \sum_{\tau=0}^{T-1} a_{n}(t+\tau).
\end{equation}
We define a Lyapunov function $L(\bU(t)) \triangleq (1/2) \sum_{n=1}^{N} U_{n}^{2}(t)$ and the $T$-slot Lyapunov drift
\[
\Delta_{T}(\bU(t)) \triangleq \expect{L(\bU(t+T) - L(\bU(t)) \mid \bU(t)},
\]
where in the last term the expectation is with respect to the randomness of the whole network in frame $T$, including the randomness of $T$. By taking square of~\eqref{eq:213} and then conditional expectation on $\bU(t)$, we can show
\begin{equation} \label{eq:214}
\begin{split}
& \Delta_{T}(\bU(t)) \leq \frac{1}{2} N (1+A_{\text{max}}^{2}) \expect{T^{2} \mid \bU(t)} \\
&- \expect{
	\sum_{n=1}^{N} U_{n}(t) \left[
		\sum_{\tau=0}^{T-1} \left(
			\mu_{n}(t+\tau) - a_{n}(t+\tau)
		\right)
	\right]
	\mid \bU(t)
}.
\end{split}
\end{equation}
Define $f(\bU(t), \theta)$ as the last term of~\eqref{eq:214}, where $\theta$ represents a scheduling policy that controls the service rates $\mu_{n}(t+\tau)$ and the frame size $T$. In the following analysis, we only consider $\theta$ in the class of $\mathsf{RandRR}$ policies, and the frame size $T$ is  the renewal period of a $\mathsf{RandRR}$ policy. By~\eqref{eq:307}, the second term of~\eqref{eq:214} is less than or equal to the constant $B_{1} \triangleq (1/2) N (1+A_{\text{max}}^{2})  \expect{(T_{\text{max}})^{2}} < \infty$. It follows that
\begin{equation} \label{eq:402}
\Delta_{T}(\bU(t)) \leq B_{1} - f(\bU(t), \theta).
\end{equation}

In $f(\bU(t), \theta)$, it is useful to consider $\theta = \mathsf{RandRR}^{*}$ and $T$ is the renewal period $T^{*}$ of $\mathsf{RandRR}^{*}$. Assume $t$ is the beginning of a renewal period. For each $n\in\{1, 2, \ldots, N\}$, because $R_{n}^{*}$ is the number of successful transmissions user $n$ receives in the renewal period $T^{*}$, we have
\[
\expect{
	\sum_{\tau=0}^{T^{*}-1} \mu_{n}(t+\tau) \mid \bU(t)
}
=
\expect{R_{n}^{*}}.
\]
Combining with~\eqref{eq:j501}, we get
\begin{equation} \label{eq:j502}
\expect{
	\sum_{\tau=0}^{T^{*}-1} \mu_{n}(t+\tau) \mid \bU(t)
} > (\lambda_{n} + \epsilon) \expect{T^{*}}.
\end{equation}
By the assumption that packet arrivals are i.i.d. over slots and independent of the current queue backlogs, we have for all $n$
\begin{equation} \label{eq:j503}
\expect{\sum_{\tau=0}^{T^{*}-1} a_{n}(t+\tau) \mid \bU(t)} = \lambda_{n} \expect{T^{*}}.
\end{equation}
Plugging~\eqref{eq:j502} and~\eqref{eq:j503} into $f(\bU(t), \mathsf{RandRR}^{*})$, we get
\begin{equation} \label{eq:403}
f(\bU(t), \mathsf{RandRR}^{*}) \geq \epsilon \expect{T^{*}} \sum_{n=1}^{N} U_{n}(t).
\end{equation}

It is also useful to consider $\theta$ as a  round robin policy $\mathsf{RR}(\bphi)$ for some $\bphi\in\Phi$. In this case frame size $T$ is the renewal period $T^{\bphi}$ of $\mathsf{RR}(\bphi)$ (note that $\mathsf{RR}(\bphi)$ is a special case of $\mathsf{RandRR}$). From Corollary~\ref{cor:201}, we have
\begin{equation} \label{eq:j509}
\expect{T^{\bphi} \mid \bU(t)} = \expect{T^{\bphi}} = \sum_{n: \phi_{n} = 1} \expect{L_{1n}^{\bphi}},
\end{equation}
where $\expect{L_{1n}^{\bphi}}$ can be expanded by~\eqref{eq:j105}. Let $t$ be the beginning of a transmission round. If channel $n$ is active, we have
\[
\expect{\sum_{\tau=0}^{T^{\bphi}-1} \mu_{n}(t+\tau) \mid \bU(t)} = \expect{L_{1n}^{\bphi}}-1,
\]
and $0$ otherwise. It follows that
\begin{equation} \label{eq:j505}
\begin{split}
&f(\bU(t), \mathsf{RR}(\bphi)) \\
&\quad =\left( \sum_{n:\phi_{n}=1} \!\! U_{n}(t) \expect{L_{1n}^{\bphi}-1}\right) \! - \expect{T^{\bphi}} \sum_{n=1}^{N} U_{n}(t)\lambda_{n} \\
& \quad \stackrel{(a)}{=} \sum_{n: \phi_{n}=1} \left[ U_{n}(t) \expect{L_{1n}^{\bphi}-1} - \expect{L_{1n}^{\bphi}} \sum_{n=1}^{N} U_{n}(t)\lambda_{n} \right],
\end{split}
\end{equation}
where (a) is by~\eqref{eq:j509} and rearranging terms.

({\bf Design of} $\mathsf{QRR}$) Given the current queue backlogs $\bU(t)$, we are interested in the policy  that maximizes $f(\bU(t), \theta)$ over all $\mathsf{RandRR}$ policies in one round of transmission. Although the $\mathsf{RandRR}$ policy space is uncountably large and thus searching for the optimal solution could be difficult, next we show that the optimal solution is a round robin policy $\mathsf{RR}(\bphi)$ for some $\bphi\in\Phi$ and can be found by maximizing $f(\bU(t), \mathsf{RR}(\bphi))$ in~\eqref{eq:j505} over $\bphi\in\Phi$. To see this, we denote by $\bphi(t)$ the binary vector associated with the $\mathsf{RR}(\bphi)$ policy that maximizes  $f(\bU(t), \mathsf{RR}(\bphi))$ over $\bphi\in\Phi$, and we have
\begin{equation} \label{eq:j506}
f(\bU(t), \mathsf{RR}(\bphi(t))) \geq f(\bU(t), \mathsf{RR}(\bphi)),\,\text{for all $\bphi\in\Phi$}.
\end{equation}
For any $\mathsf{RandRR}$ policy, conditioning on the policy $\mathsf{RR}(\bphi)$ chosen for the current round of scheduling, we have
\begin{equation} \label{eq:j507}
f(\bU(t), \mathsf{RandRR}) = \sum_{\bphi\in\Phi} \alpha_{\bphi} f(\bU(t), \mathsf{RR}(\bphi)),
\end{equation}
where $\{\alpha_{\bphi}\}_{\bphi\in\Phi}$ is the probability distribution associated with $\mathsf{RandRR}$.  By~\eqref{eq:j506}\eqref{eq:j507}, for any $\mathsf{RandRR}$ we get
\begin{equation} \label{eq:j601}
\begin{split}
f(\bU(t), \mathsf{RR}(\bphi(t))) &\geq \sum_{\bphi\in\Phi} \alpha_{\bphi} f(\bU(t), \mathsf{RR}(\bphi)) \\
&= f(\bU(t), \mathsf{RandRR}).
\end{split}
\end{equation}
We note that as long as the queue backlog vector $\bU(t)$ is not identically zero and the arrival rate vector $\blambda$ is strictly within the inner capacity bound $\Lambda_{\text{int}}$,  we get
\begin{equation} \label{eq:j602}
\begin{split}
\max_{\bphi\in\Phi} f(\bU(t), \mathsf{RR}(\bphi)) &\stackrel{(a)}{=} f(\bU(t), \mathsf{RR}(\bphi(t))) \\
&\stackrel{(b)}{\geq} f(\bU(t), \mathsf{RandRR}^{*}) \stackrel{(c)}{>} 0,
\end{split}
\end{equation}
where (a) is from the definition of $\bphi(t)$, (b) from~\eqref{eq:j601}, and (c) from~\eqref{eq:403}.

The policy $\mathsf{QRR}$ is designed to be a frame-based algorithm where at the beginning of each round we observe the current queue backlog vector $\bU(t)$, find the binary vector $\bphi(t)$ whose associated round robin policy $\mathsf{RR}(\bphi(t))$ maximizes $f(\bU(t), \mathsf{RandRR})$ over $\mathsf{RandRR}$ policies, and execute $\mathsf{RR}(\bphi(t))$ for one round of transmission. We emphasize that in every transmission round of $\mathsf{QRR}$, active channels are served in the order that the least recently used channel is served first, and the ordering may change from one round to another.


({\bf Stability Analysis})
Again, policy $\mathsf{QRR}$ comprises of a sequence of transmission rounds, where in each round $\mathsf{QRR}$ finds and executes policy $\mathsf{RR}(\bphi(t))$ for one round, and different $\bphi(t)$ may be used in different rounds. In the $k$th  round, let $T^{\mathsf{QRR}}_{k}$ denote its time duration. Define $t_{k} = \sum_{i=1}^{k} T^{\mathsf{QRR}}_{i}$ for all $k\in\N$ and note that $t_{k} - t_{k-1} = T^{\mathsf{QRR}}_{k}$. Let $t_{0} = 0$. Then for each $k\in\N$, from~\eqref{eq:402} we have
\begin{equation} \label{eq:304}
\begin{split}
\Delta_{T^{\mathsf{QRR}}_{k}}(\bU(t_{k-1})) 
&\stackrel{(a)}{\leq} B_{1} - f(\bU(t_{k-1}), \mathsf{QRR}) \\
&\stackrel{(b)}{\leq} B_{1} - f(\bU(t_{k-1}), \mathsf{RandRR}^{*}) \\
&\stackrel{(c)}{\leq} B_{1} - \epsilon\,\expect{T^{*}} \sum_{n=1}^{N} U_{n}(t_{k-1}),
\end{split}
\end{equation}
where (a) is by~\eqref{eq:402}, (b) is because $\mathsf{QRR}$ is the maximizer of $f(\bU(t_{k-1}), \mathsf{RandRR})$ over all $\mathsf{RandRR}$ policies, and (c) is by~\eqref{eq:403}. By taking expectation over $\bU(t_{k-1})$ in~\eqref{eq:304} and noting that $\expect{T^{*}} \geq 1$, for all $k\in\N$ we get
\begin{equation} \label{eq:306}
\begin{split}
&\expect{L(\bU(t_{k}))} - \expect{L(\bU(t_{k-1}))} \\
&\leq B_{1} - \epsilon\,\expect{T^{*}} \sum_{n=1}^{N} \expect{U_{n}(t_{k-1})}  \\
&\leq B_{1} - \epsilon\, \sum_{n=1}^{N} \expect{U_{n}(t_{k-1})}.
\end{split}
\end{equation}
Summing~\eqref{eq:306} over $k\in\{1, 2, \ldots, K\}$, we have
\[
\begin{split}
&\expect{L(\bU(t_{K}))} - \expect{L(\bU(t_{0}))} \\
&\qquad \leq K B_{1} - \epsilon \sum_{k=1}^{K} \sum_{n=1}^{N}\expect{U_{n}(t_{k-1})}.
\end{split}
\]
Since $\bU(t_{K})\geq \bm{0}$ entrywise and by assumption $\bU(t_{0}) = \bU(0) = \bm{0}$, we have
\begin{equation} \label{eq:j113}
\epsilon \sum_{k=1}^{K} \sum_{n=1}^{N}\expect{U_{n}(t_{k-1})} \leq K B_{1}.
\end{equation}
Dividing~\eqref{eq:j113} by $\epsilon K$ and passing $K\to\infty$, we get
\begin{equation} \label{eq:309}
\limsup_{K\to\infty} \frac{1}{K} \sum_{k=1}^{K} \sum_{n=1}^{N} \expect{U_{n}(t_{k-1})} \leq \frac{B_{1}}{\epsilon} < \infty.
\end{equation}
Equation~\eqref{eq:309} shows that the network is stable when sampled at renewal time instants $\{t_{k}\}$. Then that it is also stable when sampled over all time follows because $T_{k}^{\mathsf{QRR}}$, the renewal period of the $\mathsf{RR}(\bphi)$ policy chosen in the $k$th round of $\mathsf{QRR}$, has finite first and second moments for all $k$ (see~\eqref{eq:307}), and in every slot the number of packet arrivals to a user is bounded. These details are provided in Lemma~\ref{lem:j601}, which is proved in Appendix~\ref{pf:j601}.

\begin{lem} \label{lem:j601}
Given that
\begin{equation}\label{eq:505}
\expect{T_{k}^{\mathsf{QRR}}} \leq \expect{T_{\text{max}}}, \quad \expect{(T_{k}^{\mathsf{QRR}})^{2}} \leq \expect{(T_{\text{max}})^{2}}
\end{equation}
for all $k\in\{1, 2, \ldots\}$, packets arrivals to a user is bounded by $A_{\text{max}}$ in every slot, and the network sampled at renewal epochs $\{t_{k}\}$ is stable  from~\eqref{eq:309}, we have
\[
\limsup_{K\to\infty} \frac{1}{t_{K}} \sum_{\tau=0}^{t_{K}-1} \sum_{n=1}^{N} \expect{U_{n}(\tau)} <\infty.
\]
\end{lem}
\end{IEEEproof}

\section{} \label{pf:j601}
\begin{IEEEproof}[Proof of Lemma~\ref{lem:j601}]
In $[t_{k-1}, t_{k})$, it is easy to see for all $n\in\{1, \ldots, N\}$
\begin{equation} \label{eq:308}
U_{n}(t_{k-1}+\tau) \leq U_{n}(t_{k-1}) + \tau A_{\text{max}}, \quad 0\leq \tau < T^{\mathsf{QRR}}_{k}.
\end{equation}
Summing~\eqref{eq:308} over $\tau \in \{0, 1, \ldots, T^{\mathsf{QRR}}_{k}-1\}$, we get
\begin{equation} \label{eq:310}
\sum_{\tau=0}^{T^{\mathsf{QRR}}_{k} - 1} U_{n}(t_{k-1}+\tau) \leq T^{\mathsf{QRR}}_{k} U_{n}(t_{k-1}) + (T^{\mathsf{QRR}}_{k})^{2} A_{\text{max}}/2.
\end{equation}
Summing~\eqref{eq:310} over $k\in\{1, 2, \ldots, K\}$ and noting that $t_{K} = \sum_{k=1}^{K} T_{k}^{\mathsf{QRR}}$, we have
\begin{equation} \label{eq:j114}
\begin{split}
\sum_{\tau=0}^{t_{K}-1} U_{n}(\tau) &= \sum_{k=1}^{K} \sum_{\tau=0}^{T^{\mathsf{QRR}}_{k} - 1} U_{n}(t_{k-1}+\tau) \\
&\stackrel{(a)}{\leq} \sum_{k=1}^{K} \left[ T^{\mathsf{QRR}}_{k} U_{n}(t_{k-1}) + (T^{\mathsf{QRR}}_{k})^{2} A_{\text{max}}/2 \right],
\end{split}
\end{equation}
where (a) is by~\eqref{eq:310}. Taking expectation of~\eqref{eq:j114} and dividing it by $t_{K}$, we have
\begin{equation} \label{eq:311}
\begin{split}
&\frac{1}{t_{K}} \sum_{\tau=0}^{t_{K}-1} \expect{U_{n}(\tau)} \stackrel{(a)}{\leq} \frac{1}{K} \sum_{\tau=0}^{t_{K}-1} \expect{U_{n}(\tau)} \\
&\qquad \stackrel{(b)}{\leq} \frac{1}{K} \sum_{k=1}^{K} \expect{
T^{\mathsf{QRR}}_{k} U_{n}(t_{k-1}) + \frac{(T^{\mathsf{QRR}}_{k})^{2}A_{\text{max}}}{2}
},
\end{split}
\end{equation}
where (a) follows $t_{K}  \geq K$ and (b) is by~\eqref{eq:j114}. 
Next, we have
\begin{align}
\expect{T_{k}^{\mathsf{QRR}} U_{n}(t_{k-1})} &= \expect{\expect{T_{k}^{\mathsf{QRR}} U_{n}(t_{k-1}) \mid U_{n}(t_{k-1})}} \notag \\
&\stackrel{(a)}{\leq} \expect{\expect{T_{\text{max}} U_{n}(t_{k-1}) \mid U_{n}(t_{k-1})}} \notag \\
&=\expect{\expect{T_{\text{max}}} U_{n}(t_{k-1})} \notag \\
&=\expect{T_{\text{max}}}\expect{U_{n}(t_{k-1})},  \label{eq:j106}
\end{align}
where (a) is because $\expect{T_{k}^{\mathsf{QRR}}} \leq \expect{T_{\text{max}}}$.
Using~\eqref{eq:505}\eqref{eq:j106} to upper bound the last term of~\eqref{eq:311}, we have
\begin{equation} \label{eq:j107}
\frac{1}{t_{K}} \sum_{\tau=0}^{t_{K}-1} \expect{U_{n}(\tau)} \leq B_{2} + \expect{T_{\text{max}}} \frac{1}{K} \sum_{k=1}^{K} \expect{U_{n}(t_{k-1})}, \\
\end{equation}
where  $B_{2} \triangleq \frac{1}{2}\,\expect{(T_{\text{max}})^{2}}A_{\text{max}} <\infty$. Summing~\eqref{eq:j107} over $n\in\{1, \ldots, N\}$ and passing $K\to\infty$, we get
\begin{align*}
&\limsup_{K\to\infty} \frac{1}{t_{K}} \sum_{\tau=0}^{t_{K}-1} \sum_{n=1}^{N} \expect{U_{n}(\tau)} \\
&\quad \leq N B_{2} + \expect{T_{\text{max}}} \left(
\limsup_{K\to\infty} \frac{1}{K} \sum_{k=1}^{K} \sum_{n=1}^{N} \expect{U_{n}(t_{k-1})}
\right) \\
&\quad \stackrel{(a)}{\leq} N B_{2} + \expect{T_{\text{max}}} B_{1} / \epsilon < \infty,
\end{align*}
where (a) is by~\eqref{eq:309}. The proof is complete.
\end{IEEEproof}

\bibliographystyle{IEEEtran}
\bibliography{/Users/chihping/Desktop/bibliography/IEEEabrv,/Users/chihping/Desktop/bibliography/myabrv,/Users/chihping/Desktop/bibliography/mypaperbib}

\begin{thebibliography}{10}
\providecommand{\url}[1]{#1}
\csname url@samestyle\endcsname
\providecommand{\newblock}{\relax}
\providecommand{\bibinfo}[2]{#2}
\providecommand{\BIBentrySTDinterwordspacing}{\spaceskip=0pt\relax}
\providecommand{\BIBentryALTinterwordstretchfactor}{4}
\providecommand{\BIBentryALTinterwordspacing}{\spaceskip=\fontdimen2\font plus
\BIBentryALTinterwordstretchfactor\fontdimen3\font minus
  \fontdimen4\font\relax}
\providecommand{\BIBforeignlanguage}[2]{{%
\expandafter\ifx\csname l@#1\endcsname\relax
\typeout{** WARNING: IEEEtran.bst: No hyphenation pattern has been}%
\typeout{** loaded for the language `#1'. Using the pattern for}%
\typeout{** the default language instead.}%
\else
\language=\csname l@#1\endcsname
\fi
#2}}
\providecommand{\BIBdecl}{\relax}
\BIBdecl

\bibitem{LaN10conf-channelmemory}
C.-P. Li and M.~J. Neely, ``On achievable network capacity and
  throughput-achieving policies over markov on/off channels,'' in \emph{{IEEE}
  Int. Symp. Modeling and Optimization in Mobile, Ad Hoc, and Wireless Networks
  (WiOpt)}, Avignon, France, May 2010.

\bibitem{LaN10}
------, ``Energy-optimal scheduling with dynamic channel acquisition in
  wireless downlinks,'' \emph{{IEEE} Trans. Mobile Comput.}, vol.~9, no.~4, pp.
  527 --539, Apr. 2010.

\bibitem{WaC96}
H.~S. Wang and P.-C. Chang, ``On verifying the first-order markovian assumption
  for a rayleigh fading channel model,'' \emph{{IEEE} Trans. Veh. Technol.},
  vol.~45, no.~2, pp. 353--357, May 1996.

\bibitem{ZRM96conf}
M.~Zorzi, R.~R. Rao, and L.~B. Milstein, ``A {M}arkov model for block errors on
  fading channels,'' in \emph{Personal, Indoor and Mobile Radio Communications
  Symp. {PIMRC}}, Oct. 1996.

\bibitem{Ber05book}
D.~P. Bertsekas, \emph{Dynamic Programming and Optimal Control}, 3rd~ed.\hskip
  1em plus 0.5em minus 0.4em\relax Athena Scientific, 2005, vol.~I.

\bibitem{ZKL08}
Q.~Zhao, B.~Krishnamachari, and K.~Liu, ``On myopic sensing for multi-channel
  opportunistic access: Structure, optimality, and preformance,'' \emph{{IEEE}
  Trans. Wireless Commun.}, vol.~7, no.~12, pp. 5431--5440, Dec. 2008.

\bibitem{ALJ09}
S.~H.~A. Ahmad, M.~Liu, T.~Javidi, Q.~Zhao, and B.~Krishnamachari, ``Optimality
  of myopic sensing in multichannel opportunistic access,'' \emph{{IEEE} Trans.
  Inf. Theory}, vol.~55, no.~9, pp. 4040--4050, Sep. 2009.

\bibitem{Nee09conf}
M.~J. Neely, ``Stochastic optimization for markov modulated networks with
  application to delay constrained wireless scheduling,'' in \emph{{IEEE} Conf.
  Decision and Control (CDC)}, 2009.

\bibitem{ZaS07_2}
Q.~Zhao and A.~Swami, ``A decision-theoretic framework for opportunistic
  spectrum access,'' \emph{{IEEE} Wireless Commun. Mag.}, vol.~14, no.~4, pp.
  14--20, Aug. 2007.

\bibitem{PET07conf}
A.~Pantelidou, A.~Ephremides, and A.~L. Tits, ``Joint scheduling and routing
  for ad-hoc networks under channel state uncertainty,'' in \emph{{IEEE} Int.
  Symp. Modeling and Optimization in Mobile, Ad Hoc, and Wireless Networks
  (WiOpt)}, Apr. 2007.

\bibitem{YaS08conf}
L.~Ying and S.~Shakkottai, ``On throughput optimality with delayed
  network-state information,'' in \emph{Information Theory and Application
  Workshop (ITA)}, 2008, pp. 339--344.

\bibitem{YaS09conf}
------, ``Scheduling in mobile ad hoc networks with topology and channel-state
  uncertainty,'' in \emph{{IEEE INFOCOM}}, Rio de Janeiro, Brazil, Apr. 2009.

\bibitem{Gal96book}
R.~G. Gallager, \emph{Discrete Stochastic Processes}.\hskip 1em plus 0.5em
  minus 0.4em\relax Kluwer Academic Publishers, 1996.

\bibitem{Whi88}
P.~Whittle, ``Restless bandits: Activity allocation in a changing world,''
  \emph{{J. Appl. Probab.}}, vol.~25, pp. 287--298, 1988.

\bibitem{Git89book}
J.~C. Gittins, \emph{Multi-Armed Bandit Allocation Indices}.\hskip 1em plus
  0.5em minus 0.4em\relax New York, NY: Wiley, 1989.

\bibitem{TaE93}
L.~Tassiulas and A.~Ephremides, ``Dynamic server allocation to parallel queues
  with random varying connectivity,'' \emph{{IEEE} Trans. Inf. Theory},
  vol.~39, no.~2, pp. 466--478, Mar. 1993.

\bibitem{GNT06}
L.~Georgiadis, M.~J. Neely, and L.~Tassiulas, ``Resource allocation and
  cross-layer control in wireless networks,'' \emph{Foundations and Trends in
  Networking}, vol.~1, no.~1, 2006.

\end{thebibliography}

\end{document}